\documentclass[%
 reprint,nofootinbib,
 amsmath,amssymb,
 aps,
]{revtex4-1}

\usepackage{graphicx}
\usepackage{dcolumn}
\usepackage{bm}
\usepackage{natbib}
\usepackage{soul}
\usepackage{xcolor}

\usepackage{color}

\usepackage{subfigure}
\usepackage{amsmath}
\usepackage{enumerate}
\usepackage{caption}
\usepackage{multirow}
\usepackage{hyperref}
\usepackage{array}
\newcolumntype{P}[1]{>{\centering\arraybackslash}p{#1}}
\usepackage{fancyvrb}
\begin{document}


\title{Dependence of alpha-particle-driven Alfv\'{e}n eigenmode linear stability on device magnetic field strength and consequences for next-generation tokamaks}

\author{E. A. Tolman}
\affiliation{Plasma Science and Fusion Center, Massachusetts Institute of Technology, Cambridge, MA, USA
}
\author{N. F. Loureiro}
\affiliation{Plasma Science and Fusion Center, Massachusetts Institute of Technology, Cambridge, MA, USA
}
\author{P. Rodrigues}
\affiliation{Instituto de Plasmas e Fus$\tilde{a}$o Nuclear, Instituto Superior T\'ecnico, Universidade de Lisboa, Lisboa, Portugal}
\author{J. W. Hughes}
\affiliation{Plasma Science and Fusion Center, Massachusetts Institute of Technology, Cambridge, MA, USA
}
\author{E. S. Marmar}
\affiliation{Plasma Science and Fusion Center, Massachusetts Institute of Technology, Cambridge, MA, USA
}

\date{\today}

\begin{abstract}
Recently-proposed tokamak concepts use magnetic fields up to 12 T, far higher than in conventional devices, to reduce size and cost. Theoretical and computational study of trends in plasma behavior with increasing field strength is critical to such proposed devices. This paper considers trends in Alfv\' en eigenmode (AE) stability. Energetic particles, including alphas from D-T fusion, can destabilize AEs, possibly causing loss of alpha heat and damage to the device. AEs are sensitive to device magnetic field via the field dependence of resonances, alpha particle beta, and alpha orbit width. We describe the origin and effect of these dependences analytically and by using recently-developed numerical techniques (Rodrigues et al. 2015 Nucl. Fusion 55 083003). The work suggests high-field machines where fusion-born alphas are sub-Alfv\'enic or nearly sub-Alfv\'enic may partially cut off AE resonances, reducing growth rates of AEs and the energy of alphas interacting with them. High-field burning plasma regimes have non-negligible alpha particle beta and AE drive, but faster slowing down time, provided by high electron density, and higher field strength reduces this drive relative to low-field machines with similar power densities. The toroidal mode number of the most unstable modes will tend to be higher in high magnetic field devices. The work suggests that high magnetic field devices have unique, and potentially advantageous, AE instability properties at both low and high densities.
\end{abstract}

\maketitle


\section{Introduction}

Next-generation fusion experiments will attempt to reduce the size, cost, and complexity of fusion power plants. One way to do this is to increase volumetric fusion power density, which scales as \cite{wesson2004tokamaks, sorbom2015arc,hughes2018access}
\begin{equation}
\label{eq:highb}
\frac{P_f}{V_p} \propto \left< p \right>^2 \propto \frac{\beta_N^2 I_p^2 B_0^2}{ a^2},
\end{equation} 
where $P_f$ is the fusion power, $V_p$ is the plasma volume, $\left< p\right>$ is the volume-averaged plasma pressure, $B_0$ is the toroidal magnetic field, $I_p$ is the plasma current, $a$ is the plasma minor radius, and $\beta_N$ is the normalized beta, a parameter limited by stability considerations to roughly 3 in conventional tokamaks \cite{troyon1984mhd}.  Increasing magnetic field at fixed $\beta_N$ and fixed or increasing $I_p$ allows improvement in fusion power density. The recent development of high-temperature superconducting (HTS) technology \cite{fietz2005high} has allowed the exploitation of this trend in conceptual designs for next-generation fusion experiments with large fusion energy gain factor $Q \equiv P_{ fusion}/P_{ heat}$, with $ P_{ fusion}$ the fusion power and $P_{ heat}$ the amount of heating power supplied,  that remain compact and cost-efficient by using magnetic fields as strong as $B_0 = 9.2$ T  \cite{sorbom2015arc}. The recently-announced SPARC concept will have an even larger field of around 12 T \cite{psfcreport}, as might machines proposed in the future. Most modern tokamaks operate at on-axis magnetic fields below 4 T, and the extensively-analyzed ITER baseline scenario has $B_0  = 5.3$ T \cite{shimada2007overview}. Thus, developing theoretical and computational understanding of trends in tokamak behavior when $B_0$ increases beyond these values is of critical importance for the new era of very-high-field HTS machines.  
 
 This paper focuses on developing this understanding in one area  of particular significance to high-magnetic-field experiments which hope to demonstrate net energy gain: the stability behavior of Alfv\'{e}n eigenmodes (AEs). 
AEs are shear Alfv\'{e}n waves that exist in tokamaks as discrete modes and which can be driven unstable by energetic particles, leading to increased energetic particle transport. In a future D-T device, AEs could lead to both loss of alpha power needed to heat the plasma and to damage to device walls \cite{heidbrink2008basic}. The majority of the present-day understanding of AEs results from analysis of AEs driven by heating mechanisms such as neutral beam injectors (NBI) and ion cyclotron resonance heating  (ICRH). This experience has led to understanding which allows extrapolation to future experiments aimed at producing net fusion energy, where significant AE drive will come both from heating sources and from alpha particles.   \cite{heidbrink2008basic}
 
In particular, this understanding can be used to project the magnetic field dependence of alpha and heating drive of AEs in future devices. The trend in heating drive is strongly dependent on the specifics of the heating system selected for a given experiment, but a basic estimate of trends with magnetic field may still be conducted (see Section~\ref{sec:heating}), showing that the overall dependence of AE drive from heating sources on magnetic field strength is expected to be weak (when neglecting the effect of choice of heating type). Because of  the strong dependence on the choice of heating system during machine design, this source of AE drive is not the focus of this paper. Rather, this paper focuses on alpha drive for AEs, one of the most novel elements of AE behavior in future devices.

Alpha drive of AEs is also expected to show particularly interesting trends with increasing magnetic field due to the sensitivity of its physics to device magnetic field. For example, alpha particles are born with a small energy dispersion around 3.5 MeV, i.e., with a birth velocity $v_{\alpha 0 } \approx 1.3 \times 10^7 \, \rm m/s$, independent of device magnetic field. The most important resonances of the toroidal Alfv\'{e}n eigenmode (TAE) with plasma particles occur at plasma particle velocity parallel to the magnetic field $v_{\parallel} \sim v_{A0}$ and $v_{\parallel} \sim v_{A0}/3$, with $v_{A0} = B_0 / \sqrt{\mu_0 \rho_0 }$ the on-axis Alfv\'{e}n speed and $\rho_0$ the on-axis value of the plasma mass density, $\rho \equiv \sum_i n_i m_i$ with ion species of number density $n_i$ and mass $m_i$. Other AE modes (i.e., the ellipticity-induced EAE and the noncircularity-induced NAE) also have their primary resonances at speeds of order the Alfv\'{e}n speed. We may define a parameter 
\begin{equation}
\label{eq:lambda}
\lambda \equiv \frac{v_{A0}}{v_{\alpha 0}}
\end{equation}
characterizing the relationship between a given device's Alfv\'{e}n speed and the alpha particle birth velocity. If device $\beta$ and $\beta_N$ are kept roughly constant as magnetic field increases, and Greenwald fraction, defined in equation ~\eqref{eq:fgw}, is kept constant or decreasing, $\lambda$ will increase with magnetic field. When $\lambda$ surpasses 1, alpha particles will never interact with the higher $v_\parallel \sim v_{A0}$ TAE resonance, reducing TAE drive. Alphas will, however, still interact with lower resonances. The alpha particle beta, $\beta_\alpha$,  is also strongly dependent on magnetic field due to the dependence of the absolute values of bulk plasma density and temperature on magnetic field, as this paper will demonstrate. This behavior introduces $\lambda$ dependencies even below $\lambda = 1$, as will other less significant dependencies introduced later in the paper.

Several previous studies of AE stability in next-generation devices have been published \cite{gorelenkov2003study,rodrigues2016sensitivity,yang2017linear,chen2010linear,rodrigues2015systematic,figueiredo2016comprehensive}. AE physics is highly sensitive to a variety of device parameters, and these works have focused on describing the dependence of stability on several of them, including ion depletion factor, ion temperature, plasma $\beta$ \cite{gorelenkov2003study}, core temperature gradient \cite{jaun2000stability,gorelenkov2003study,figueiredo2016comprehensive} and $q$ profile \cite{rodrigues2016sensitivity,yang2017linear}. Toroidal magnetic field dependence (encoded in $\lambda$) has not been a focus, perhaps because of the limited magnetic fields at which many future devices were proposed to operate. The AE stability of some high-field concepts, including IGNITOR \cite{gorelenkov2003study}, has been considered to some degree, but these studies have not critically examined the way in which the stability features of these devices is caused by high field, and thus are not sufficient to provide an overall understanding of high-field trends. Several tools for understanding AE stability, including a code suite used in this paper, were also not available at the time such studies were conducted.

Recently-proposed high-magnetic field devices have reinvigorated interest in this area. These machines will likely have the ability to explore high $\lambda$  regimes, and even reach the $\lambda \gtrsim 1$ regime.  The ARC reactor's baseline scenario \cite{sorbom2015arc}, with $n_{i,0} = 1.8 \times 10^{20} \, \rm m^{-3}$ and $B_0 = 9.2$ T, gives, assuming equal concentrations of deuterium and tritium, $\lambda = 0.73$; small modifications in the operating parameters could lead to $\lambda \gtrsim 1$.  The recently-announced SPARC tokamak concept \cite{psfcreport} could also explore this regime. Thus, a robust understanding of the dependence on magnetic field of the physics of AE stability, which includes but is not limited to the loss of resonances, is of critical importance. Of particular interest is understanding the unique challenges and advantages high magnetic field devices may have when dealing with AE behavior, especially if this knowledge can be exploited in the design process.

The goal of this paper is to develop and codify this understanding for AE linear stability. To do so, it draws from substantial analytical and computational machinery developed over the past several decades to study the behavior of AEs. Specifically, the paper consists of three sections. In the Section~\ref{sec:deps}, we describe the dominant trend with magnetic field of tokamak parameters which influence AE stability. Then, in Section~\ref{sec:analytical}, we build on the existing literature of theoretical studies of AE stability to suggest how AE behavior is likely to change as device magnetic field increases, focusing on mode structure, the strength of the linear growth rate, the energy of the alpha particles interacting with the mode, and the toroidal mode number of the most unstable mode. Section~\ref{sec:heating} provides an extension of these arguments to heating drive. Then, in Section~\ref{sec:computational}, we use a suite of codes \cite{rodrigues2015systematic} to study numerically the AE behavior of a realistic tokamak equilibrium which is scaled through magnetic field in a way that represents the dominant magnetic field trends identified in Section~\ref{sec:deps}.

Together, these sections create a picture through which to evaluate the AE stability properties of the high magnetic field path. As the paper will demonstrate, high magnetic field machines will tend to operate with higher $\beta_\alpha$ than low field machines, and this $\beta_\alpha$ will increase the alpha particle drive to the AEs in the machine. However, as the conclusion (Section~\ref{sec:pow}) will explain, high magnetic field devices require less $\beta_\alpha$ to produce a given power density, suggesting that when economic considerations are taken into account, the trend in $\beta_\alpha$ with field becomes an advantage.  Furthermore, high magnetic field devices are capable of partially or fully cutting off resonances between AEs and alpha particles, reducing AE drive and the energy of alpha particles interacting with the AE.
Meanwhile, the mode structure of AEs will not change as a result of device magnetic field, while the toroidal mode number of the most unstable mode will increase with magnetic field, which could influence its interaction with magnetic field ripple.

\section{Dependence of device parameters on magnetic field}
\label{sec:deps}
At a given magnetic field, a tokamak may operate with a wide range of plasma parameters (density, current, pressure, etc.). The value of each of these parameters will affect AE stability \cite{rodrigues2016sensitivity,yang2017linear,gorelenkov2003study}, and thus, at a given magnetic field, a tokamak will have significant flexibility to shape its AE stability properties. However, each of these parameters will also have a dominant trend with device magnetic field strength; for example, as equation~\eqref{eq:highb} indicates, higher magnetic field devices will tend to operate at higher absolute pressures. In this case, any effects of higher device pressure on AE stability properties are also indirect effects of higher magnetic field operation; the same is true for other trends in device parameters with magnetic field. This section aims to lay out these overall trends.  Sections~\ref{sec:analytical} and~\ref{sec:computational} will then consider the effect of these trends, in addition to direct effects of magnetic field strength. The overall trends discussed in the remainder of this section are summarized in Table~\ref{tab:tab1} for convenience.
\begin{table}
\centering
\begin{tabular}{ | m{2.5cm} | m{2.5cm}| }
\hline
Quantity& Scaling with $B_0$\\ 
\hline
\hline
$p$ & $\sim B_0^2$  \\ 
\hline
$f_{GW}$ & $\sim B_0^{-\xi}$\\
\hline
$I_p$ & $\sim B_0$\\
\hline
$n_e$ & $ \sim B_0^{1-\xi}$\\
\hline
$T_i, \, T_e$ & dependent on $p_\alpha$\\
\hline
$q$ & constant\\
\hline
\end{tabular}
\caption{Dominant trends in tokamak quantities with magnetic field (i.e., those that result from keeping $\beta$ and $\beta_N$  fixed and letting $f_{gw}$ vary with magnetic field as $f_{gw} \sim B_0 ^{- \xi}$).}
\label{tab:tab1}
\end{table}

\subsection{Equilibrium}
\label{sec:mageq}
The structure and real frequency of the AEs present in a given tokamak will be determined by its MHD equilibrium. 
These equilibria will be solutions of the Grad-Shafranov equation \cite{grad1958hydromagnetic,shafranov1960equilibrium,lust1957axialsymmetrische}, 
\begin{equation}
\label{eq:gs}
R \frac{\partial}{\partial R} \left(\frac{1}{R} \frac{\partial \psi}{\partial R} \right) + \frac{\partial^2 \psi}{\partial Z^2} =  - \mu_0 R^2 \frac{dp}{d \psi} - \frac{1}{2} \frac{d F^2}{d \psi},
\end{equation}
where $R$ is the distance to the torus' axis, $Z$ is the height above the midplane, $\psi$ is the poloidal flux, $p$ is the plasma pressure, and the parameter $F\equiv R B_\phi$ is the product of major radius and toroidal magnetic field, which is related to the poloidal current in the plasma and the toroidal field coils. 

This equation has a symmetry that allows it to be scaled through size and magnetic field strength without affecting the shape of equilibrium functions or device figures of merit.  
In particular, the equation is symmetric under the transformation 
\begin{equation}
\label{eq:col}
\begin{cases}
R \rightarrow cR
\\
Z \rightarrow cZ
\\
\psi \rightarrow \psi_N \psi
\\
 p \left( \psi \right) \rightarrow \frac{\psi_N^2}{c^4} p \left( \frac{\psi}{\psi_N} \right)
\\
F^2 \left( \psi\right) \rightarrow \frac{\psi_N^2}{c^2} F^2 \left( \frac{\psi}{\psi_N} \right),
\end{cases}
\end{equation} 
with $c$ and $\psi_N$ numerical scaling constants. These transformations also mean that the toroidal magnetic field, the current density, and the overall plasma current change according to
\begin{equation}
\label{eq:quant}
\begin{cases}
B_{\phi} \rightarrow \frac{\psi_N}{c^2}B_{\phi}
\\
J_{\phi} \rightarrow \frac{\psi_N}{c^3} J_{\phi}
\\
I_p \rightarrow \frac{\psi_N}{c^3} c^2 I_p \sim \frac{\psi_N}{c} I_p,
\end{cases}
\end{equation}
and thus that 
\begin{equation}
\label{eq:param}
\begin{cases}
\beta \equiv \frac{p}{B_0^2/ \left( 2 \mu_0 \right)}  \rightarrow \beta
\\
\beta_N  \equiv \frac{\beta a B_0}{I_p} \rightarrow \beta_N
\\
q\left(\psi \right) \equiv  \frac{F \left( \psi \right)}{2 \pi} \int \frac{dl}{R^2 B_p} \rightarrow q\left(\psi / \psi_N\right),
\end{cases}
\end{equation}
with $B_p$ the poloidal magnetic field.
These transformations alter device size and magnetic field strength without altering device figures of merit $\beta$ and $\beta_N$ or the \textit{shape} of the pressure, current, and $q$ profiles. However, plasma current and pressure will increase as magnetic field strength does, though the maintenance of constant $\beta_N$ ensures that the resulting configuration is consistent with MHD stability limits. Because this transformation changes field without changing figures of merit and increases pressure and current in the way frequently implied when considering the high field approach to fusion, we take it as defining the overall trend in equilibrium quantities with field.

As pressure increases, plasma density and temperature will necessarily increase. As discussed in the following section, the maintenance of constant Greenwald fraction will ensure that the density does not rise beyond limits allowed by the current.  We note that increasing plasma temperature is frequently observed to change plasma wall interaction in a way that increases the impurity accumulation in the core of the plasma \cite{hughes2018access,neu2007plasma}, such that increasing temperature sufficiently to keep a constant $\beta$ may be contingent on wall material or on having a core plasma which is good at removing impurities (such as an I-mode). This trend is, however, considered to be of secondary importance to  the overall improved confinement at high field leading to increasing pressure at roughly constant $\beta$.

\subsection{Profiles}
\label{sec:prof}
Next, we consider the magnetic field dependence of the density and temperature profiles of the species present in the plasma. In particular, we consider the magnetic field trends of the profiles of electrons, DT ions, alpha particles, and helium ash.
\subsubsection{Electron and DT ion density}
The electrons  in a tokamak can be characterized by the profiles $n_e \left(\psi \right)$ and $T_e \left( \psi \right)$; each of the ion species may be characterized by the profiles $n_i \left( \psi \right)$ and $T_i \left( \psi \right)$. In this paper, we assume that the dominant ion species are deuterium and tritium, with $n_D\left(\psi \right) = n_T \left(\psi \right)= n_e \left(\psi \right) /2$. We choose not to consider the  effect of ion dilution $\sigma = \left(n_D + n_T \right)/n_e$, as this parameter is not expected to exhibit a strong scaling with on-axis device magnetic field; a discussion of its influence on AE linear stability may be found in \cite{gorelenkov2003study}.

The dominant scaling of density with magnetic field is determined through the Greenwald fraction \cite{greenwald1988new}, which is defined as
\begin{equation}
\label{eq:fgw}
f_{GW} \equiv \frac{\bar{n}_e \left[ 10^{20} \, \textrm{m}^{-3}\right]}{I_p[\textrm{MA}]/ \pi a[\textrm{m}]^2},
\end{equation}
with $a$ the device minor radius and $\bar{n}_e$ the line-averaged $n_e$.
Disruptivity limits this parameter to \cite{greenwald1988new}
\begin{equation}
f_{GW} \le 1.
\end{equation}
When a tokamak operates in L-mode or I-mode it has great freedom to vary the Greenwald fraction. In this case, since the constant-figure-of-merit scaling given by \eqref{eq:col} through \eqref{eq:param} gives plasma current increasing linearly with magnetic field for fixed device size, density will likewise tend to increase linearly with device magnetic field for a given choice of $f_{GW}$.

However, operation in H-mode is somewhat more complicated. In some types of H-mode, like the EDA H-mode routinely observed on Alcator C-Mod, the density pedestal height, which strongly correlates with the overall density, is found to be determined by shot parameters at the time of the L-H transition and insensitive to attempts to modify the density pedestal height after the H-mode transition \cite{hughes2002observations,hughes2013pedestal,hughes2007edge}.  For EDA H-mode, the Greenwald fraction defined with the density pedestal height has been experimentally found \cite{tolman2018influence} to scale as
\begin{multline}
\label{eq:tolscale}
f_{GW,ped} \equiv \frac{n_{e,ped}}{I_p / \left( \pi a^2 \right)}
\\
 = \left(0.83\right) \times \left(\bar{n}_{e,L} \left[ 10^{20}\, \textrm{m}^{-3} \right]\right)^{0.22} \times \left( B_0 \left[ \textrm{T} \right] \right)^{-0.64 }.
\end{multline}
Here, $n_{e,ped}$ is the density at the top of the density pedestal and $\bar{n}_{e,L}$ is the line-averaged density of the L-mode that created the H-mode, which represents how strongly the shot was fueled.  Because the density pedestal height is strongly correlated with the overall density, this indicates that the shot Greenwald fraction decreases strongly with magnetic field. 

Some types of H-mode can be fueled during the H-mode phase, but these H-modes exhibit an H-mode density limit, which corresponds to a maximum line-averaged density, below the Greenwald limit, at which the plasma back transitions to an L-mode \cite{greenwald2002density}. An H-mode can thus vary its Greenwald fraction below this limit, but not above it; if the tokamak seeks to maximize its density, the density will be determined by this limit. Regressions over a very limited magnetic field range (roughly 1 T to 3 T) have found that the Greenwald fraction corresponding to this H-mode density limit declines sharply with magnetic field, but the full magnetic field behavior has not been explored \cite{borrass2004recent,bernert2014h}.

These behaviors may be represented together through the statement
\begin{equation}
f_{GW} \sim B_0^{-\xi},
\end{equation}
where $\xi$ may be $0$ or a finite number depending on if operating regimes with a dominant scaling of Greenwald fraction with magnetic field are to be considered.  From the scaling of current with magnetic field ~\eqref{eq:quant} $I_p \sim B_0$, we can thus state
\begin{equation}
n_e \sim B_0^{1-\xi}.
\end{equation}

\subsubsection{Electron and DT ion temperature}
The pressure of a plasma composed of electrons, deuterium and tritium ions, and alpha particles is given by 
\begin{multline} 
p\left( \psi \right) = 
n_\alpha \left( \psi \right) T_\alpha \left( \psi \right) + n_e \left(\psi \right) T_e \left( \psi \right) + n_D \left(\psi \right) T_D \left(\psi \right)
\\
 + n_T \left( \psi \right) T_T \left(\psi \right).
\end{multline} 
Care must be taken in defining the alpha temperature because the alpha particles are represented by a slowing down distribution, not a Maxwellian (see Section~\ref{sec:alphas}). We have that
\begin{equation}
n_\alpha T_\alpha = \frac{1}{3} m_\alpha \int v^2 f_\alpha d^3 \vec{v},
\end{equation}
which should be evaluated for a slowing down distribution~\eqref{eq:origsd}.
This yields a temperature roughly given by
\begin{equation}
\label{eq:alphtemp}
T_\alpha \sim E_{\alpha 0} / 6  \sim 580 \,  \rm keV.
\end{equation} 
We consider regimes with $n_D = n_T = n_e/2$; let us further take that $T_D = T_T = T_e \equiv T$. As will be discussed in Section~\ref{sec:alphas}, alpha density is determined as a function of the local density and temperature, such that 
\begin{equation}
n_\alpha \left( \psi \right) = n_e\left(\psi \right) g\left(T\left( \psi \right)\right),
\end{equation}
where the function $g$ may be found from~\eqref{eq:alphdens}.  Thus, the plasma temperature is related to the plasma pressure via
\begin{multline}
p\left(\psi \right) 
\\ =  n_e\left( \psi \right) \left[  g\left(T\left(\psi\right) \right) T_\alpha+ 2T \left( \psi \right) \right] 
\equiv n_e \left(\psi\right) f\left(T\left(\psi \right)\right).
\end{multline}
From this statement, we may solve for temperature as
\begin{equation}
\label{eq:temp}
T \left(\psi \right) = f^{-1} \left( \frac{p\left(\psi \right)}{n_e\left(\psi\right)}\right),
\end{equation}
which yields the temperature necessary to obtain the given pressure at a given density. Since pressure will grow with magnetic field squared at constant $\beta$, and density will grow linearly or slower than linearly with magnetic field, temperature will, over regimes of interest, grow with magnetic field strength.
Given the complicated nature of the formula for the cross section,~\eqref{eq:temp} is not straightforward to evaluate analytically, but may be easily found numerically.
For cases where $n_\alpha T_\alpha \ll n_e T_e$, equation~\eqref{eq:temp} reduces to the familiar $T = p/\left(2n_e\right)$.

\subsubsection{Alpha particles}
\label{sec:alphas}
Alpha particles born with a source rate $S$ in a deuterium-tritium plasma are described by the slowing down distribution
\begin{equation}
\label{eq:origsd}
f_\alpha \left(v, \psi \right) = \frac{S \epsilon_0^2 m_{ equiv} m_\alpha}{4 n_e e^4 \ln \Lambda} \left( \frac{1}{1 + v^3/v_{crit}^3}\right), \, v<v_{\alpha 0},
\end{equation}
with $v_{\alpha 0}$ the alpha particle birth velocity, $v_{\alpha 0 } = 1.3 \times 10^7 \,  \rm  m/s$,
\begin{equation}
v_{crit} = 3^{1/3}  \left( \pi /2 \right)^{1/6} \left[ T_e / \left(m_e^{1/3} m_{equiv}^{2/3} \right)\right]^{1/2},
\end{equation}
and 
\begin{equation}
m_{equiv}  \equiv \frac{2m_D m_T}{m_D + m_T}. 
\end{equation}
A derivation of this is presented in Appendix~\ref{sec:slowing}. Sometimes, the distribution is modified to account for the small energy dispersion around 3.5 MeV; in this case, the distribution is as given in equation~\eqref{eq:aspackslow}. For D-T fusion alphas, the source rate is given by 
\begin{equation}
\label{eq:S}
S = n_D n_T \left< \sigma v \right>_{DT}.
\end{equation}
Thus, the alpha particle density is given by

\begin{multline}
\label{eq:alphdens}
n_{\alpha}\left( n_e \left(\psi \right), T\left( \psi \right) \right) =
\\
 \frac{ n_D n_T \left< \sigma v \right>_{DT}\epsilon_0^2 m_{equiv} m_\alpha}{4 n_e e^4 \ln \Lambda} \int_0^{v_{\alpha 0}} \left( \frac{4 \pi v^2}{1 + v^3/v_{crit}^3}\right) dv 
\\
= \frac{\pi n_e \left( \psi \right) \left< \sigma v \right>_{DT} \left(T \left(\psi \right) \right)\epsilon_0^2 m_{equiv} m_\alpha v_{crit}\left( T \left(\psi \right)\right)^3}{12 e^4 \ln \Lambda}
\\
\times \ln \left(1 + \frac{v_{\alpha 0}^3}{v_{crit}\left(T\left(\psi \right) \right)^3} \right).
\end{multline}
We take $\ln \Lambda = 17$; the fusion reactivity $\left< \sigma v \right>_{DT} \left(T\right)$ may be parameterized as follows:

\begin{equation}
\left< \sigma v \right>_{DT} = 6.4 \times 10^{-20}m^{3} s^{-1} x^{-5/6}y^2 \exp{\left( -3 x^{1/3} y\right)},
\end{equation}
where, with $T$ expressed in keV, 
\begin{multline}
x =1- 
\\
\frac{\left(1.5\times 10^{-2} \right)T+ \left(4.6\times 10^{-3}\right)T^2 - \left(1.1\times 10^{-4}\right)T^3}{1 + \left(7.5 \times 10^{-2} \right)T+ \left( 1.4 \times 10^{-2}\right)T^2 + \left(1.4 \times 10^{-5} \right)T^3} ,
\end{multline}
and 
\begin{equation}
y = \frac{6.7}{T^{1/3}}.
\end{equation}

\begin{figure*}[!ht]
\centering
\includegraphics[width=1.3\columnwidth]{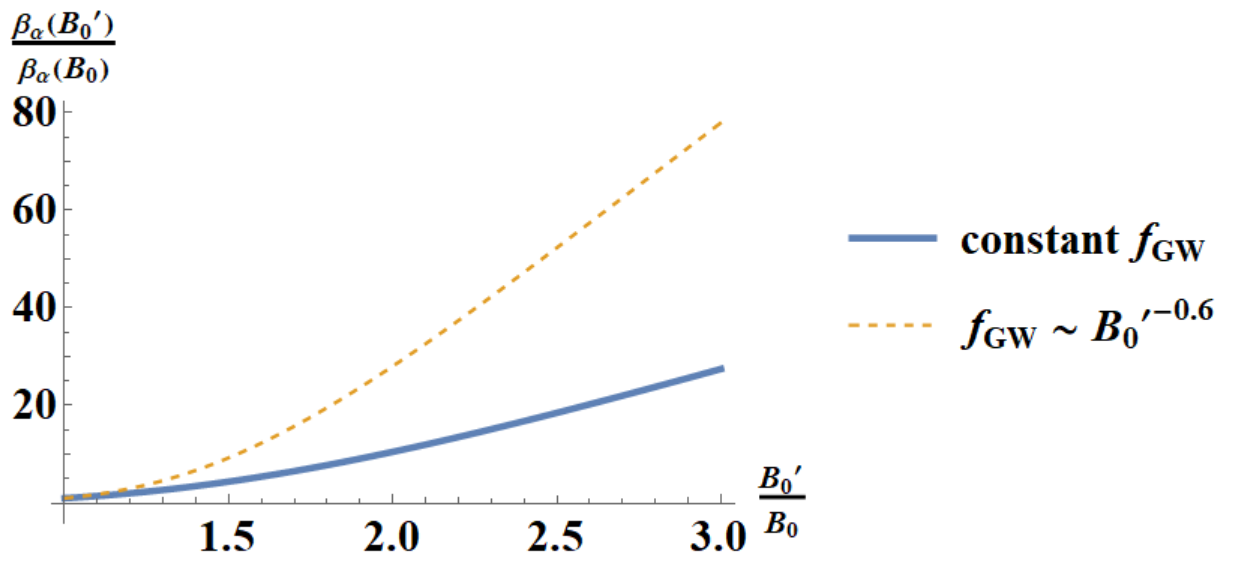}
\caption{$\beta_\alpha$ as a function of magnetic field $B_0 '$ relative to a reference magnetic field $B_0$ at which the plasma temperature is 5 keV, given by~\eqref{eq:alphbet}, for both a case where $f_{GW}$ is taken to be constant with magnetic field and a case where Greenwald fraction decreases with magnetic field (the exponent characterizing the decrease is chosen to match that in~\eqref{eq:tolscale}). Temperature increases with field; over the range presented here, the temperature of the decreasing Greenwald fraction case ranges from 5 to 24.6 keV, while that of the constant Greenwald fraction case ranges from 5 to 13.9 keV.}
\label{fig:betaalpha}
\end{figure*}

The alpha contribution to the AE growth rate increases strongly with $\beta_\alpha \equiv 2 \mu_0 n_\alpha T_\alpha /B_0^2$ (see Section~\ref{sec:analytical}). Thus, we wish to obtain an understanding of the dominant scaling of this quantity with magnetic field strength.
We consider the ratio of $\beta_\alpha$ at one field with respect to that at another field,
\begin{multline}
\label{eq:alphbet}
\frac{\beta_\alpha \left(B_0' \right)}{ \beta_\alpha \left(B_0  \right)}  = \frac{B_0^2 T_\alpha \left(B_0' \right)n_\alpha \left( B_0 '\right)}{ B_0'^2 T_\alpha \left(B_0\right) n_\alpha \left( B_0 \right) }
\\
 \approx \left( \frac{B_0 }{B_0'} \right)^{1+\xi} \frac{\left< \sigma v \right>_{DT} \left(f^{-1} \left(\frac{p\left(B_0\right)}{n_e\left(B_0\right)} \left( \frac{B_0'}{B_0} \right)^{1+\xi} \right) \right)}{\left< \sigma v \right>_{DT}\left(f^{-1} \left(\frac{p\left(B_0\right)}{n_e\left(B_0\right)}\right)\right)}
 \\
 \times \left(\frac{f^{-1} \left(\frac{p\left(B_0\right)}{n_e\left(B_0\right)} \left( \frac{B_0'}{B_0} \right)^{1+\xi} \right) }{f^{-1} \left(\frac{p\left(B_0\right)}{n_e\left(B_0\right)}\right)}\right)^{3/2}
 .
\end{multline}
In deriving this expression, we have taken $T_\alpha \sim E_{\alpha 0} / 6$ (see equation~\eqref{eq:alphtemp}), used~\eqref{eq:alphdens}, and recalled that $p \sim B_0^2$ while $n_e \sim B_0^{1-\xi}$, with $\xi$ representing any dominant scaling of Greenwald fraction.  Plotting this quantity for both constant and decreasing Greenwald fraction cases gives Figure~\ref{fig:betaalpha}, which shows the growth in $\beta_\alpha$ over fields that correspond to an increase in temperature from 5 to 13.9 keV in the case of constant $f_{GW}$ and 5 to 24.6 keV in the case of decreasing $f_{GW}$.
This plot demonstrates that $\beta_\alpha$ increases with magnetic field over plasma temperatures of interest, indicating that higher magnetic field machines will tend to have higher $\beta_\alpha$ than lower magnetic field machines.\footnote{While equation~\eqref{eq:alphbet} is difficult to parse due to its correct and detailed treatment of alpha pressure and fusion reactivity, the general behavior may be understood through the following simplified argument. Because the temperature of slowing-down alphas is approximately constant, $\beta_\alpha$ will scale approximately as $\beta_\alpha \sim \frac{n_\alpha}{B_0^2}$; over the region where $\left<\sigma v \right>_{DT} \sim T^2$, this will be given by $\beta_\alpha \sim \frac{n_e^2 T^2}{B_0^2 n_e /T^{3/2}}$, where the factor of density over temperature to the three halves in the denominator represents how quickly alphas slow down. Simplifying this expression for $n_e \sim B_0$, $T\sim B_0$ (as would be the case for neglecting alpha pressure) gives $\beta_\alpha \sim B_0^{2.5}$.}

Also of note is that decreasing Greenwald fraction at fixed pressure (as the dotted orange line demonstrates) will in general increase $\beta_\alpha$. This is due to two effects.
 First, decreasing density increases the amount of time it takes alpha particles to slow down on the background plasma, such that the overall alpha pressure is higher.  Second, decreasing plasma density at fixed pressure, such that temperature increases, will increase the fusion reactivity.
\subsubsection{He ash and other impurities}
Helium ash and other impurities will build up in the plasma at a rate dependent on how quickly the ash is transported out of the plasma and pumped out.  Given the time-dependent nature of this quantity and its complicated nature, we choose to neglect it.  Its contribution to AE instability is also expected to be small--it is of roughly equivalent temperature to the deuterium and tritium ions, but is of significantly smaller density \cite{rodrigues2015systematic} .

\section{Analytical stability trends}
\label{sec:analytical}
The previous section introduced the dominant trends with magnetic field strength of tokamak parameters with particular importance to AE stability. In this section, we consider the effects that these trends actually have on AE stability. To do this, we use results from analytic study of AEs, which provides insight into the computational results that follow in Section~\ref{sec:computational}.
\subsection{Mode structure}
\label{sec:struc}
Prior to considering how growth rates of the AEs present in high magnetic field devices will tend to change with increasing magnetic field strength, we begin by showing that the structure of the modes present will not tend to change with magnetic field.
The transformations~\eqref{eq:col} leave the shapes of the profiles unchanged,  so the shapes and frequency normalized to on-axis Alfv\'{e}n frequency (which scales with magnetic field divided by radius) of the eigenmodes existing in equilibria related by this transformation will be the same.  This may be shown explicitly from the MHD eigenmode equation, which determines the mode real frequency and structure:
\begin{equation}
\label{eq:eigenmode}
-\rho \omega^2 \vec{\xi}_\perp = \vec{F}_\perp \left( \vec{\xi}_\perp \right)  \equiv \vec{J}_1 \times \vec{B}_{eq} + \vec{J}_{eq} \times \vec{B}_1 + \nabla \left( \vec{\xi}_\perp  \cdot \nabla p \right), 
\end{equation}
where $\omega$ is the mode real frequency, $\vec{\xi}_\perp$ is the perturbed $\vec{E} \times \vec{B}$ displacement, $\vec{E}_{1 \perp} \equiv i \omega \vec{\xi}_\perp \times \vec{B}_{eq}$, $\vec{F}_\perp$ is the ideal MHD force operator, $\vec{J}_1$ and $\vec{B}_1$ are the perturbed current and magnetic field, respectively, and the subscript $_{eq}$ denotes equilibrium quantities. Scaling the mode frequency to the on-axis Alfv\'{e}n frequency $\omega_{A0} \equiv B_0 / \left( R_0 \sqrt{\mu_0 \rho_0}\right)$ gives
\begin{equation}
- \frac{B_0^2}{R_0^2 \mu_0 }  \frac{\rho}{\rho_0}  \left(\frac{\omega}{\omega_{A0}}\right)^2 \vec{\xi}_\perp = \vec{J}_1 \times \vec{B}_{eq} + \vec{J}_{eq} \times \vec{B}_1 + \nabla \left( \vec{\xi}_\perp  \cdot \nabla p  \right).
\end{equation}
Scaling this equation according to the transformations \eqref{eq:col} through \eqref{eq:param} shows that the shape of $\vec{\xi}_\perp$ and the value of $\omega /\omega_{A0}$ will be the same as long as the shape of the density profile, $\rho/ \rho_0$, is unaltered.

\subsection{Value of growth rate}
\label{sec:strengthgamma}
A given tokamak will have a variety of AEs which have the potential to become excited. The symmetry discussed in Section~\ref{sec:struc} indicates that the set of existing modes with the potential to become excited does not depend on magnetic field.  However, the linear growth rate of each of these modes will depend strongly on magnetic field.

Total linear AE growth or damping rate is determined by the contributions from each of the species in the plasma interacting with the AE:
\begin{equation}
\label{eq:sum}
\frac{\gamma}{\omega} = \frac{\gamma_\alpha}{\omega} + \sum_j \frac{\gamma_j}{\omega},
\end{equation}
where the mode growth rate $\gamma$, expressed as a ratio of the mode frequency $\omega$, is given as a sum of the contribution due to alpha particles and due to thermal plasma species. Each of these contributions may be dependent on magnetic field.

To obtain a straightforward, simplified view of AE stability, many analytical calculations of these contributions ignore the finite width of AE modes, considering them to have one discrete radial localization, and consider only fully passing particles. In this section, we make both of these assumptions in order to understand the basic trends in AE stability with device field strength. These assumptions are not present in the numerical treatment that forms Section~\ref{sec:computational}; one of the main effects of the loss of these assumptions is that discrete AE resonances become spread out in phase space. However, even with the more precise numerical treatment, the framework developed in this section serves well to understand overall behavior.

Analytically, the contribution of alpha particles to the AE growth rate is estimated to be \cite{betti1992stability}
\begin{equation}
\label{eq:alphagrow}
\frac{\gamma_\alpha}{\omega} \sim -q_{AE}^2 \beta_\alpha \left( G^\alpha \left( \lambda \right) - n q_{AE} \delta_\alpha H^\alpha \left(\lambda \right)  \right),
\end{equation}
with $\lambda$ the parameter defined in~\eqref{eq:lambda}, $n$ the toroidal mode number, $q_{AE}$ the value of $q$ at the location of the mode, $\delta_\alpha \equiv -\left(2/3\right) r_{L \theta} \left(dp_\alpha /dr \right)/ p_\alpha$ the ratio of the alpha particle birth poloidal gyroradius, $r_{L \theta} \equiv v_{\alpha 0}/ \Omega_{\theta \alpha}$ with $\Omega_{\theta \alpha}\equiv q_\alpha B_p /m_\alpha$,  to the pressure scale length. The function $G^\alpha\left(\lambda \right)$ represents damping of the mode on the alpha particles due to the gradient of their distribution function with respect to energy, $\frac{\partial f_\alpha}{\partial E}$, at points where the mode is resonant with the alpha particle population. The function $H^\alpha \left(\lambda \right)$ represents growth of the mode due to the spatial gradient of the alpha particle distribution function, $\frac{\partial f_\alpha}{\partial \psi}$, at the points where the mode is resonant.  The location of the resonant points depends on the type of mode. For example, the TAE has its primary resonances at alpha particle velocities $v_\parallel = v_{A0}$ and $v_\parallel = v_{A0}/3$, while the EAE has its primary resonance at $v_\parallel = v_{A0}/2$.  

The alpha particle energy distribution function will be given by a slowing down distribution,~\eqref{eq:origsd} (or~\eqref{eq:aspackslow} if the finite birth energy dispersion around 3.5 MeV is included); the shape of the distribution function in $\psi$ will depend on the equilibrium. The slowing down distribution is pictured in Figure~\ref{fig:slow}.

Importantly, this distribution function includes only particles with energy below roughly $E_{\alpha 0} = 3.5 \,  \rm MeV$, the alpha particle birth velocity (some particles will have slightly more energy than 3.5 MeV due to retained energy from the collision that caused the fusion); furthermore, more alpha particles are at lower energies than at higher energies.  This means that if the energy corresponding to one of the resonances of an AE lies above 3.5 MeV, that resonance will not contribute to the growth rate of the AE, reducing the overall drive to the AE.  For the TAE, the first loss of a resonance will occur at $\lambda \equiv v_{A0}/v_{\alpha 0} = 1$.  However, since more alpha particles are at lower energies than at higher energies, the loss of this resonance is expected to remove only a fraction of the overall AE drive.

Meanwhile, the factor $\beta_\alpha$ in~\eqref{eq:alphagrow} multiplies the part of the expression dependent on resonances. As discussed in Section~\ref{sec:alphas}, $\beta_\alpha$ will tend to increase with magnetic field, such that we expect this factor to cause a strong increase in AE drive with magnetic field. This increase, combined with any decrease due to loss of resonances, will determine the overall evolution of growth rate as a function of magnetic field. 

We must also consider the damping contributions to the AE.  These result from the interaction of the AE with bulk thermal species and are given by \cite{betti1992stability}
\begin{equation}
\label{eq:thermaldamp}
\frac{\gamma_j}{\omega} \sim -q_{AE}^2 \beta_j G^T\left(\lambda_T\right).
\end{equation} 
Here, $G^T\left(\lambda_T\right)$ represents damping due to the energy gradient of the thermal species distribution, $\partial f_j/ \partial E$, at the points where the mode is resonant with the particles; $\lambda_T = v_{A0}/v_{Tj}$ with $v_{Tj}$ the thermal velocity of the species in question.  Since bulk species have a Maxwellian distribution, there will be no complete loss of a resonance when the magnetic field becomes too strong, as there was with a slowing down distribution, which has a sharp cutoff at the birth energy. Likewise, $\beta_j$ for bulk species tends to remain constant with magnetic field. The damping of AEs will thus not display strong trends with magnetic field, and instead will change gradually only due to the changing value of  $\partial f_j/ \partial E$ at the resonance location. 
\subsection{Energy of alpha particles interacting with mode}
\label{sec:energyinteract}
\begin{figure*}[!ht]
\centering
\subfigure[$\,$ Slowing down distribution] {\label{fig:slow}\includegraphics[width=.9\columnwidth]{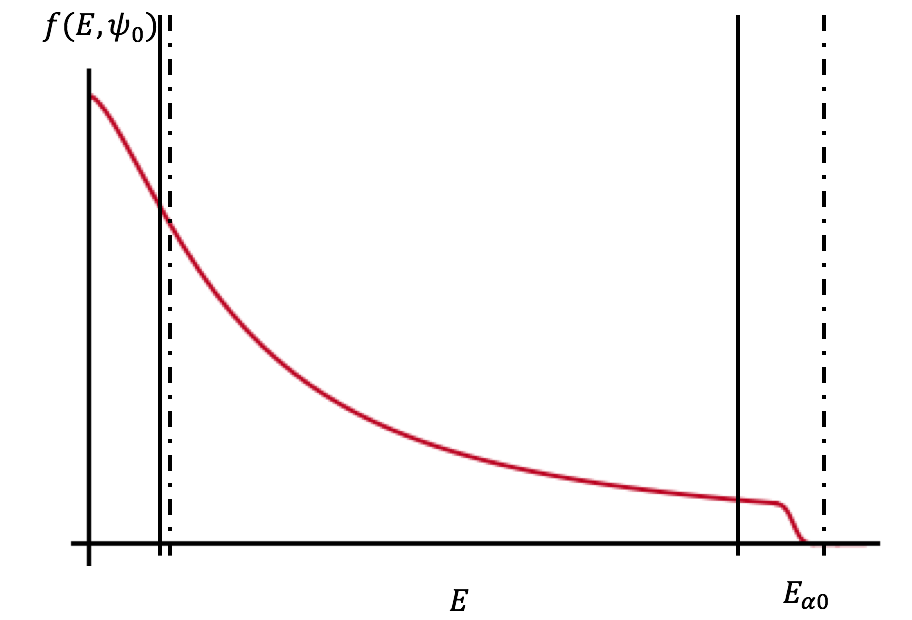}}
\subfigure[ $\,$ Slowing down distribution weighted by energy]{\label{fig:weightedsd}\includegraphics[width=.9\columnwidth]{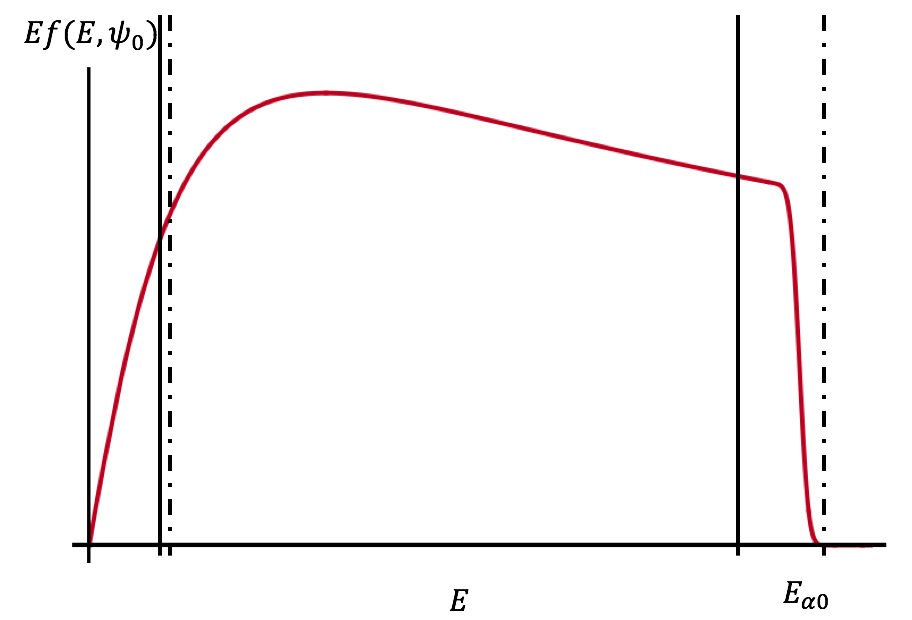}}
\caption{The energy distribution of fusion alpha particles; primary TAE resonance locations are indicated schematically for $\lambda <1$ with solid lines and for $\lambda > 1$ with dotted lines.}
\end{figure*}

As discussed in Section~\ref{sec:strengthgamma}, the AE interacts with alpha particles through a set of resonances. In the case of the TAE, the higher energy resonance at $v_\parallel = v_{A0}$ will be cut off in machines with high enough field. The lower resonance $v_\parallel = v_{A0}/3$ will still provide substantial drive to the mode.  However, the growth rate of a mode is not the only important result of an analysis of linear AE physics. Also important is the energy of the alpha particles that are able to resonate with the mode, as these will be the alpha particles which have the potential to be transported out of the core by AEs. The loss of the upper resonance means that only lower-energy particles can be affected by the mode. This will reduce the loss of higher energy alphas, which have more chance of damaging the machine and also carry more heating power than lower-energy alphas. A visual representation of this effect is given in Figure~\ref{fig:weightedsd}, which shows the slowing down distribution weighted by energy, showing that although more particles exist at lower energy, the energy content is roughly evenly distributed between lower and higher energy, such that cutting off a higher resonance will greatly reduce the overall energy of the alpha particles that interact with the AE.

\subsection{Toroidal mode number of most unstable mode}
\label{sec:nmax}
While all AEs in the tokamak will, regardless of mode number, experience the general trends described in Sections~\ref{sec:strengthgamma} through~\ref{sec:energyinteract}, there will be another effect that modulates this trend and leads to a change in which mode is the most unstable at a given field.

The width of an AE may be estimated by taking~\eqref{eq:eigenmode} and solving for a mode of the form $\vec{\xi}_\perp = \exp\left[ i \left( n \phi - \omega t \right)\right] \sum_m \vec{\xi}_{\perp, m} \left( r \right) \exp{\left(-i m \theta\right)}$. Stating the result to zeroth order in inverse aspect ratio  gives \cite{breizman1995energetic,rosenbluth1992continuum,pinches1996nonlinear}
\begin{equation}
\frac{d}{dr} \left( \frac{\omega^2}{v_A^2} - k_{ \parallel m}^2\right)  \frac{d \vec{\xi}_{\perp,m }}{dr} - \frac{m^2}{r^2} \left( \frac{\omega^2}{v_A^2} - k_{ \parallel m}^2\right) \vec{\xi}{\perp, m} = 0,
\end{equation}
with $k_{\parallel m} \equiv \left(1/R\right) \left[ n - m/q\left(r\right)\right]$. 
Balancing the terms in this equation gives that the mode width 
\begin{equation}
\label{eq:modew}
\Delta_{ m} \sim r_{m} / m, 
\end{equation}
with $r_{m}$ the radial location of the mode.  For TAEs, $q \left(r_m \right) = \left( m - 1/2 \right) /n$, while for EAEs $q \left(r_m \right) = \left( m - 1 \right) /n$, and similar relationships hold for other types of AE.  Thus~\eqref{eq:modew} may be approximated by 
\begin{equation}
\label{eq:modew2}
\Delta_{m} \sim r_{m} / n q \left(r_{m}\right).
\end{equation}
Meanwhile, the width of an alpha particle orbit for a particle with $v_\parallel \gg v_\perp$ is given by 
\begin{equation}
\label{eq:modewidth}
\Delta_{ orbit} \sim \frac{q v_\parallel}{\Omega_\alpha},
\end{equation}
with $\Omega_\alpha$ the alpha particle cyclotron frequency. 

As suggested by equation~\eqref{eq:alphagrow}, TAE growth rate will roughly tend to increase with increasing $n$ when orbit width effects do not have to be considered; the same trend is true for other types of AE. However, once the alpha particle orbit width exceeds the mode width, the growth rate will decrease with increasing $n$ because the coupling between the mode and the particle is less effective \cite{heidbrink2003alfven,breizman1995energetic,berk1992finite,fu1992excitation}. The result is that the maximally excited modes are those with toroidal mode numbers given roughly by $\Delta_{ m} \sim \Delta_{ orbit} $, or
\begin{equation} 
\label{eq:nmax}
n_{\max} \sim \frac{r \Omega_\alpha}{v_\parallel q^2}.
\end{equation}
We note that experimental evidence for a similar scaling has been presented in  \cite{heidbrink2002alpha}.
Because $\Omega_\alpha = q_\alpha B_0/m_\alpha$, while under the trends outlined in Table~\ref{tab:tab1}, 
\begin{equation}
v_{A0} \sim \sqrt{B_0} B_0^{\xi/2},
\end{equation}
 the toroidal mode number of the most unstable mode will increase with magnetic field except in the case of very significant decrease in Greenwald fraction with magnetic field.  This increase will be affected as well by the changing parallel velocity of the resonant particles. For the case of the TAE,  lower magnetic field machines with $\lambda < 1$ will have resonant particles with $v_\parallel \sim v_{A0}$ and $v_\parallel \sim v_{A0}/3$, while machines with $\lambda >1$ will have contributions exclusively from particles with  $v_\parallel \sim v_{A0}/3$, further enhancing the increase in $n$.

\section{Extension of results to heating drive for AEs}
\label{sec:heating}
The main focus of this paper is alpha particle drive of AEs. However, a rough extension of this analysis may be made to heating source drive for AEs.  While this will give a very broad understanding of how heating drive of AEs will scale with magnetic field, the analysis must be repeated for specific experimental concepts, because details will be contingent on the heating method selected.

A rough estimate for the amount of heating power a device will require to operate is given by the L-H threshold power, which represents the amount of loss power required to obtain enhanced performance (a similar analysis could be conducted with the L-I threshold power).  The most recent scaling for this threshold is \cite{martin2008power}
\begin{equation}
P_{th} = 0.0488e^{\pm 0.057}n_{e20}^{0.7171 \pm 0.035} B_0^{0.803 \pm 0.032}S^{0.941 \pm 0.019}.
\end{equation}
Here, $P_{th}$ is in units of MW, $n_{e20}$ is the line-average density in $10^{20}$ m$^{-3}$, $B_0$ is in T, and $S$ is the plasma surface area in m$^2.$ Recent work has supported the ability of this scaling to reproduce trends at high field \cite{tolman2018influence,schmidtmayr2018investigation}. We assume that this power is provided exclusively by ICRH for concreteness. In addition, this is the primary heating mechanism used by SPARC and has several attractive attributes for reactors, including ability to operate at high density and smaller size relative to NBI \cite{psfcreport}. Future work should examine other heating sources.

The beta resulting from a given ICRH power is  given roughly by \cite{mantsinen2000new}
\begin{equation}
\beta_{ICRH} \sim \frac{\frac{P_{ICRH} T_e^{3/2}}{n_e}}{B_0^2},
\end{equation}
which balances the ICRH power supplying the fast ions against the speed at which slowing down occurs.  Now, taking $P_{ICRH} \sim P_{th}$ , neglecting changes in plasma surface area, and taking $n_e \propto B_0$, $T_e \propto B_0$ (the actual scaling of temperature with magnetic field will be weaker due to the contribution of alpha pressure), we find that 
\begin{equation}
\beta_{ICRH} \sim B_0^{0.02}.
\end{equation}
(Inclusion of the weaker temperature growth with magnetic field would reduce the exponent somewhat.) This scaling indicates that the beta from ICRH will not have a strong trend with field.

Since fast ions resulting from ICRH will not have a sharp, discrete birth velocity like alpha particles do, a sharp loss of a resonance, like that which occurs with alpha particle drive, will not exist. The effect of changes in resonance location on heating drive of AEs is expected to be gradual and of less importance than the overall trend in beta. Thus, we expect that, overall, the magnetic field a device operates at will not have a strong effect on the heating drive to its AEs.

\section{Systematic AE linear stability analysis of SPARC-sized tokamak scanned through magnetic field}
\label{sec:computational}

To confirm the trends suggested in Section~\ref{sec:analytical}, we computationally study the AE mode structure and linear stability of an artificial tokamak equilibrium which is scanned through magnetic field in the sense described in Section~\ref{sec:deps}. The use of this computational scan provides more tangible demonstration of  magnetic field trends than analytics can provide, and it also allows the inclusion of effects not included in analytical treatments, including finite mode width and a variety of particle orbits. The computational study is conducted on TAEs for specificity; other types of AEs are expected to show similar trends, but with different resonance locations.   
Because AE stability properties vary strongly with the factors mentioned in the introduction, including core temperature gradient, ion depletion, $q$-profile, and plasma beta, this scan is not intended to represent the stability of any particular machine.  However, for concreteness, some parameters are selected to be similar to those that might be used in the SPARC concept, as this is a very-high-magnetic field tokamak concept currently undergoing development which is similar in size to DIII-D and ASDEX Upgrade, two currently operating tokamaks.

This process is conducted using a workflow developed in \cite{rodrigues2015systematic}.  In the workflow, the tokamak MHD equilibrium is first refined using the equilibrium code HELENA \cite{huysmans1991cp90}.  Then, the ideal MHD stability code MISHKA \cite{mikhailovskii1997optimization} is used to scan frequency space to find frequency eigenvalues which correspond to well-resolved eigenmodes.  Eigenmodes that intersect the Alfv\'{e}n continuum are also discarded; these modes experience significant damping and are unlikely to be the most important in a device; furthermore, existing estimates of continuum damping rates do not suggest magnetic field dependence \cite{pinches2015energetic}. The growth rate of the remaining eigenmodes is evaluated using the drift-kinetic code CASTOR-K \cite{borba1999castor}.

The codes used in this workflow have been successfully used to model JET experiments \cite{nabais2015castor, nabais2018tae}; in addition, they have been used to study an ITER baseline scenario,  and may be successfully compared with other studies of the same scenario \cite{rodrigues2015systematic,pinches2015energetic,lauber2015local}.  The codes include the physics necessary to observe the magnetic field trends outlined in Section~\ref{sec:analytical}, including finite alpha orbit width and finite mode width. However, they do not include certain effects, including most notably the damping of the modes that results from finite gyroradius terms $k_\perp \rho_i$ and $k_\perp \rho_\alpha$.   These effects are better treated by other codes \cite{holod2009electromagnetic,rodrigues2015systematic,figueiredo2016comprehensive}, but such codes are too computationally intensive to use for the frequency-space scans of multiple tokamak configurations needed here, and are thus not implemented in a workflow like that developed in \cite{rodrigues2015systematic}. 

Finite gyroradius corrections first become important through the alpha gyroradius, when $k_\perp \rho_\alpha \sim 1$ (further discussion of finite ion gyroradius effects may be found in~\cite{rodrigues2015systematic,figueiredo2016comprehensive,pinches2015energetic}). From~\eqref{eq:modew2}, this condition becomes
\begin{equation}
\left(n q/r_{m} \right) \rho_\alpha \sim 1.
\end{equation}
Estimating $r_{m} \sim 0.5 a$, with $a$ the tokamak minor radius, $q \sim 1$, and $\rho_\alpha = \left(a/90\right) / \left(B_0/ 10 \,  \rm T\right)$, where we have taken $a \sim 0.5 \, \rm  m$, and estimated a typical $\rho_\alpha$ by taking the average perpendicular component of the alpha thermal velocity corresponding to~\eqref{eq:alphtemp}, gives that these corrections become important for toroidal mode numbers near
\begin{equation}
 n \sim 45 \left( \frac{B_0}{10 \,  \rm T} \right).
\end{equation}
While the effect of finite alpha gyroradius corrections cannot be computationally studied using the methods in this paper, we analytically comment on their likely effect in Section~\ref{sec:probeff}. If in the future a computationally efficient suite of codes is developed with both the abilities of our present suite of codes and the ability to include finite gyroradius terms, the study in this section could be repeated, though we do not expect significant changes for the modes we focus on at SPARC-like fields.

\begin{figure*}[!ht]
\centering
\subfigure[ $\,$ Density and temperature profiles at 7T]{\label{fig:profs} \includegraphics[width=.9\columnwidth]{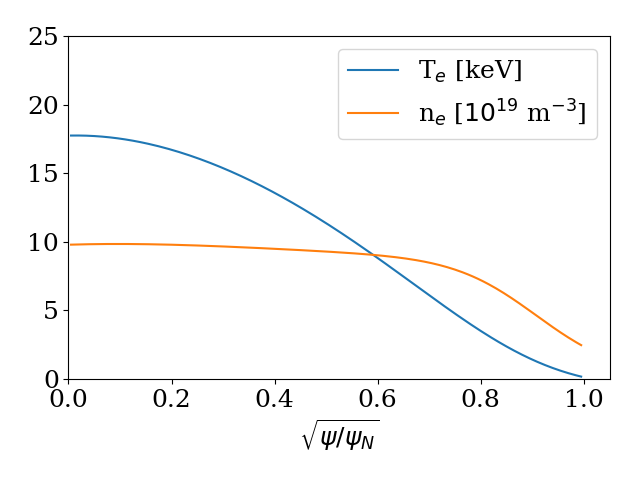}}
\subfigure[ $\,$ Density and temperature profiles at 10T]{\label{fig:profs} \includegraphics[width=.9\columnwidth]{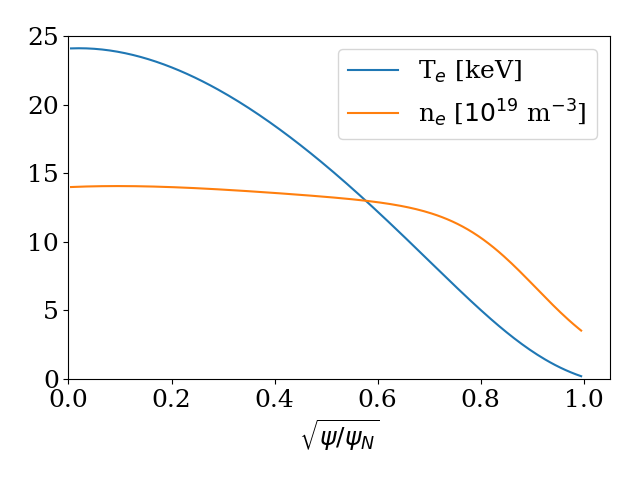}}
\subfigure[ $\,$ $q$ profile (same at all fields)]{\label{fig:qprof}\includegraphics[width=.9\columnwidth]{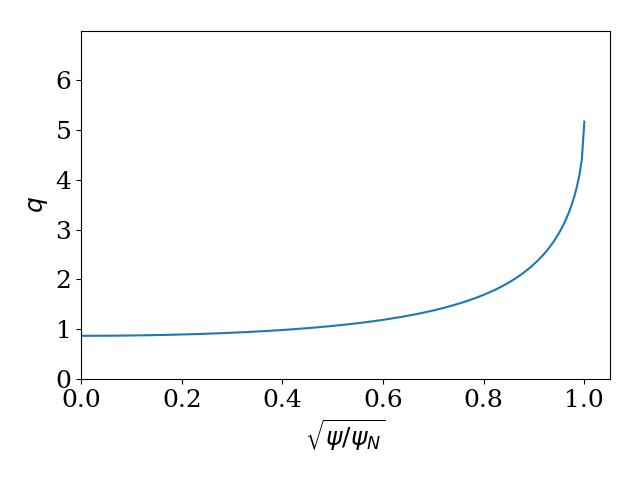}}
\subfigure[ $\,$ Alpha particle profile at 7 T] {\label{fig:alphas}\includegraphics[width=.9\columnwidth]{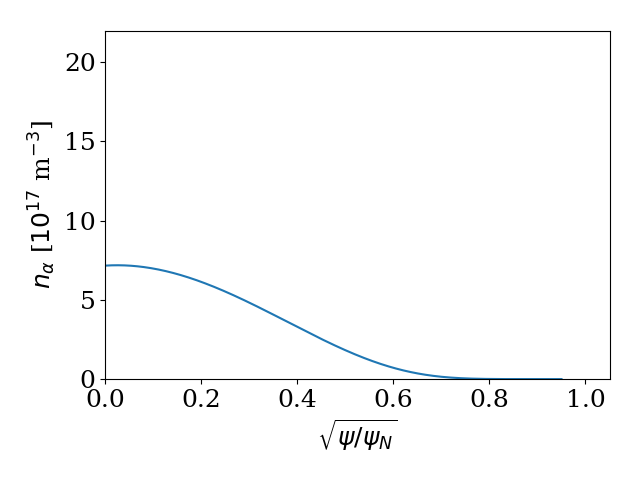}}
\subfigure[ $\,$ Alpha particle profile at 10 T] {\label{fig:alphas}\includegraphics[width=.9\columnwidth]{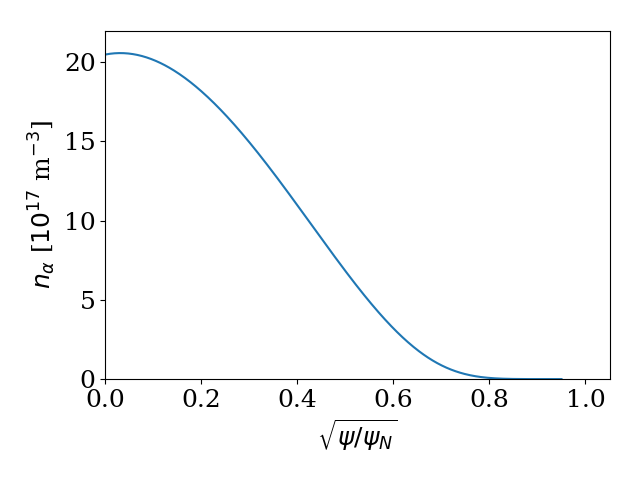}}
\caption{Profiles at two different fields of the equilibrium which is scanned through different values of the toroidal magnetic field and evaluated for stability.}
\label{fig:equilib}
\end{figure*}

To conduct a study of the stability of AEs as a function of magnetic field, a model set of tokamak profiles is needed. The trends presented in this section are found to be robust to equilibrium choice, but for the purpose of this paper, we use an equilibrium based on Alcator C-Mod discharge 1150923014 at 0.724 seconds.  This is a 5.7 T I-mode with minor radius $a \sim 0.2$ m and $q_{95} \sim 3.5$, slightly greater than a potential value for $q_{95}$ in a SPARC-like fusion device \cite{psfcreport}. The I-mode regime has favorable properties for the operation of compact, high magnetic field, net energy machines \cite{sorbom2015arc,hubbard2017physics}. The pressure and current profiles for our study are obtained by taking a kinetic equilibrium reconstruction performed on this shot and scaling to a magnetic field of choice and to a major radius $R_0$ of 1.65 m, intended to be roughly the same size as the SPARC concept \cite{psfcreport}, using transformations presented in Section~\ref{sec:mageq}, such that the shape of the pressure profile, the shape and the values of the $q$ profile, and the values of $\beta$ and $\beta_N$ are the same in the reconstruction of the C-Mod shot and in the model tokamak. The shape of the density profile and the Greenwald fraction are taken to be the same as in C-Mod shot.  Some figures of merit for the shot are summarized in Table~\ref{tab:tab3}.
\begin{table}
\centering
\begin{tabular}{ | m{2.5cm} | m{2.5cm}| }
\hline
Quantity& Value\\ 
\hline
\hline
$\beta$ &0.75\%  \\ 
\hline
$f_{GW}$ & $0.18$\\
\hline
$\beta_{N}$ & $0.80 $\\
\hline
$q_{95}$ & $ 3.5$\\
\hline
\end{tabular}
\caption{Figures of merit for the equilibrium used for systematic scan.}
\label{tab:tab3}
\end{table}
The profiles for this scenario, evaluated at 7 T and 10 T, are shown in Figure~\ref{fig:equilib}.

The alpha particle density is then calculated as though the tokamak were operating with a 50-50 deuterium and tritium blend. That is, equation~\eqref{eq:alphdens} is evaluated at each point in the plasma with the local temperature and density, with $n_D = n_T = n_e/2$. The temperature is calculated using~\eqref{eq:temp}; that is, it is set such that the total pressure resulting from the bulk plasma and the alpha particles self-consistently produced by that bulk plasma equals the pressure used to calculate the equilibrium. 
In the workflow \cite{rodrigues2015systematic}, the alpha particle slowing down distribution is taken to be a version of~\eqref{eq:origsd} modified to include the effect of a small energy dispersion about the initial birth energy:
\begin{equation}
\label{eq:aspackslow}
f_{\alpha} \left(s,E \right) = n_\alpha \left(s \right) \sqrt{E} f_{sd} \left(E\right) / \int_0^\infty dE \sqrt{E} f_{sd} \left(E \right),
\end{equation}
with
\begin{equation}
f_{sd} \left(E \right) = \frac{1}{E^{3/2} + E_c^{3/2}} \textrm{erfc}  \left( \frac{E-E_0}{\Delta_E} \right),
\end{equation}
with $E_0 = 3.5$ MeV, $\Delta_E = 50$ keV, and $E_c$ taken to be a constant, $E_c = 730$ keV.  The alpha particle density at 7 T and 10 T is shown in Figure~\ref{fig:equilib}.

\subsection{Mode structure}
\label{sec:eigen}

The set of TAEs in this equilibrium is found using MISHKA, following the frequency scan and filtering process described in \cite{rodrigues2015systematic}.
This process finds 77 TAEs; the set of eigenmodes is the same regardless of the magnetic field strength, in agreement with Section~\ref{sec:struc}. Four TAEs are selected in this section to demonstrate typical TAE growth rate evolution with field; these are pictured in Figure~\ref{fig:modesused}.
Note the noticeably smaller width of the higher-$n$ modes with respect to the lower-$n$ modes, which confirms the expected trend from~\eqref{eq:modewidth}. 
\subsection{Value of growth rate}
\begin{figure*}[!ht]
\centering
\subfigure[ $\, n = 18$, $\omega / \omega_{A0} = 0.59$ TAE]{\label{fig:firsttae} \includegraphics[width=.9\columnwidth]{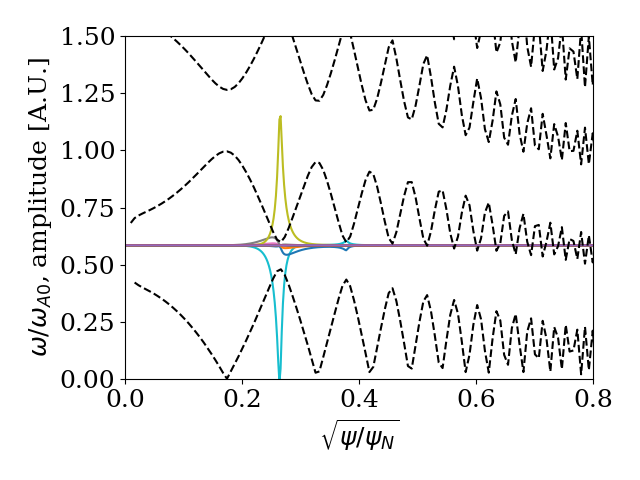}}
\subfigure[ $ \, n = 21$, $\omega / \omega_{A0} = 0.58$ TAE]{\label{fig:secondtae}\includegraphics[width=.9\columnwidth]{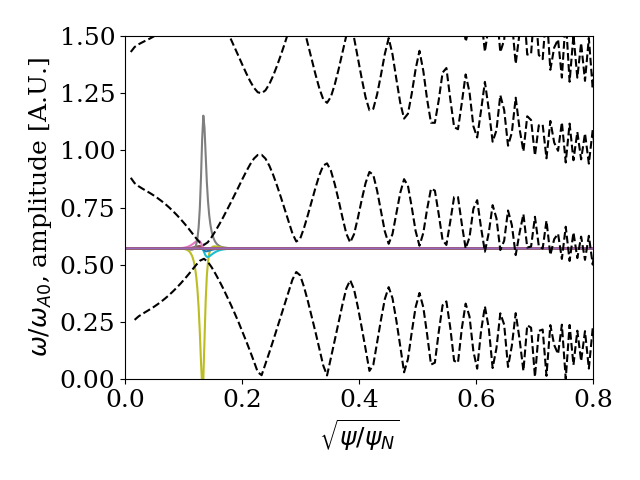}}
\subfigure[ $\, n = 22$, $\omega / \omega_{A0} = 0.45$ TAE] {\label{fig:thirdtae}\includegraphics[width=.9\columnwidth]{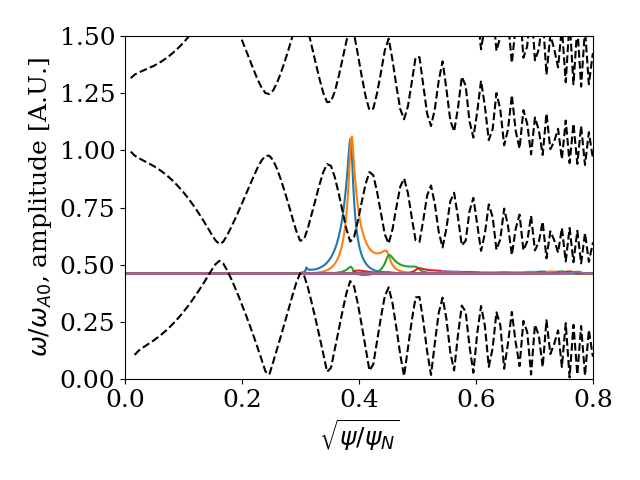}}
\subfigure[ $\,n = 33$, $\omega / \omega_{A0} = 0.46$ TAE] {\label{fig:fourthtae}\includegraphics[width=.9\columnwidth]{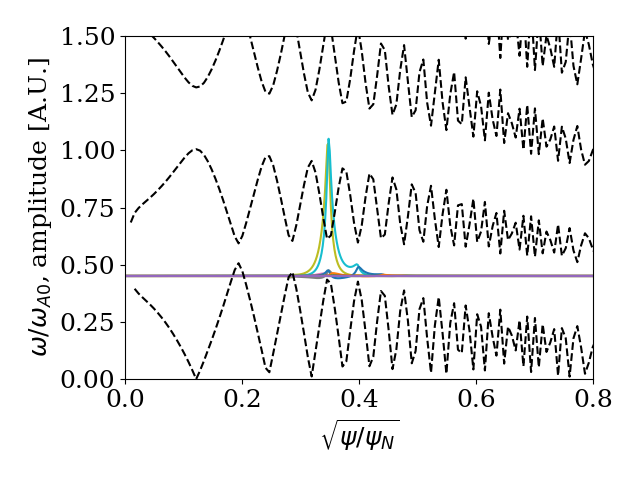}}
\caption{Structure of modes whose stability properties are examined in detail in this section, plotted with the continuum spectrum (black dashed line) at the corresponding value of $n$. Note that because MISHKA, the code used to compute the continuum spectrum, uses incompressible MHD, the lowest, BAE gap is not included. Different colored lines represent different values of the poloidal mode number $m$. For each mode, 16 consecutive poloidal harmonics are pictured. The initial poloidal mode number is, clockwise from top left, $m_0 = 8$, $m_0 = 11$, $m_0 = 23$, and $m_0 = 21$.}
\label{fig:modesused}
\end{figure*}

In this section, we consider how the contribution of each of the plasma species to the growth rate evolves with magnetic field. That is, we seek to verify and expand on the trends identified in~\ref{sec:strengthgamma} for each of the species contributing to~\eqref{eq:sum}. This process is conducted with the modes shown in Figure~\ref{fig:modesused}.  

The modes are selected to represent two different classes of TAEs; ~\ref{fig:firsttae} and~\ref{fig:secondtae} are modes located in the core of the device with small radial extent and a simple structure;~\ref{fig:thirdtae} and~\ref{fig:fourthtae} are more mid-radius modes of more complex structure. Within each of these two categories, modes of lower and higher $n$ values are selected. CASTOR-K is used to calculate the contribution of each plasma species to the overall growth rate of the mode. In all cases, the damping of the modes on electrons is found to be negligibly small in comparison to damping on deuterium and tritium, so it is neglected. Though only four modes are shown in this section of the paper, many more modes, and many other equilibrium configurations, were examined in the course of this study; all showed similar trends.

\begin{figure*}[!ht]
\centering
\subfigure[$\,$Alpha particle contribution to growth rate, normalized to 10 T on-axis Alfv\'{e}n frequency]{\label{fig:algam} \includegraphics[width=.9\columnwidth]{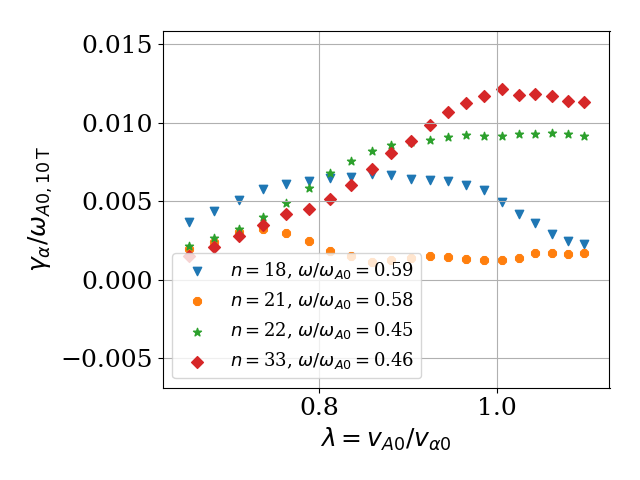}}
\subfigure[$\,$Alpha particle contribution to growth rate, normalized to mode frequency and to amount of $\beta_\alpha$]{\label{fig:normalgam}\includegraphics[width=.9\columnwidth]{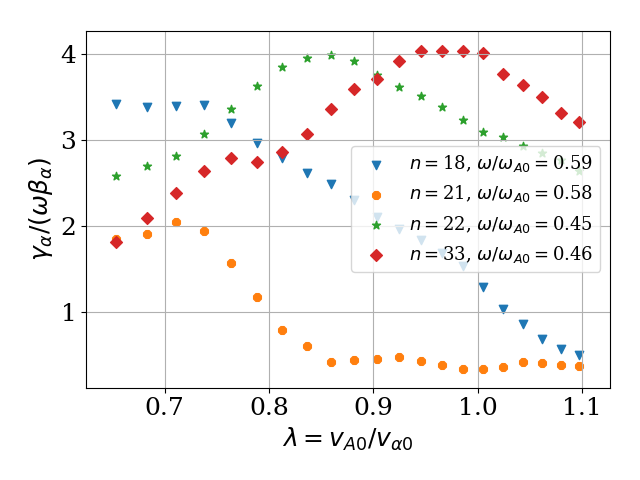}}
\subfigure[$\,$Deuterium contribution to growth rate, normalized to 10 T on-axis Alfv\'{e}n frequency (tritium contribution displays same trends)] {\label{fig:dgam}\includegraphics[width=.9\columnwidth]{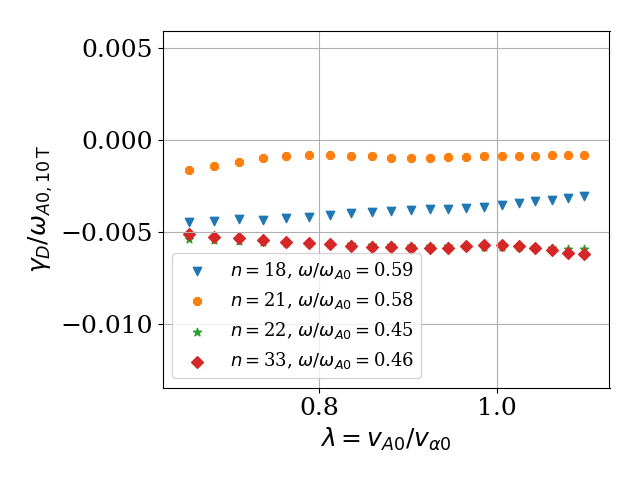}}
\subfigure[$\,$Overall growth rate, normalized to  10 T on-axis Alfv\'{e}n frequency, including contributions from alpha particles, deuterium, and tritium (electron contribution is found to be negligible)] {\label{fig:overallgam}\includegraphics[width=.9\columnwidth]{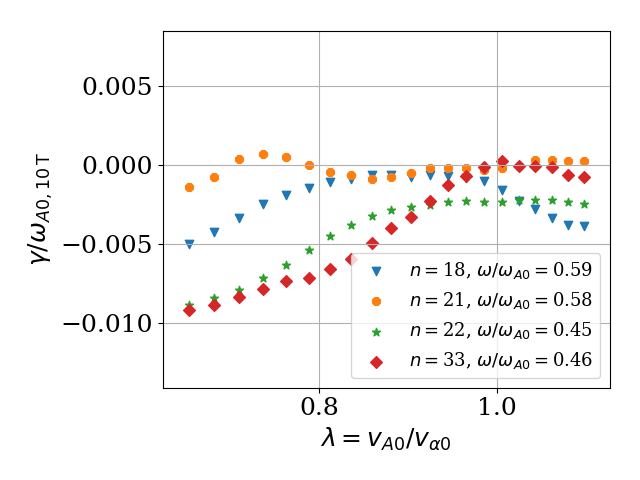}}
\caption{Evolution of relevant AE growth rates as device magnetic field is scanned. The lowest magnetic field plotted is 5.5 T, while the highest magnetic field plotted is 15.5 T. $\lambda = 1$ occurs at 13 T. }
\label{fig:trends}
\end{figure*}

The results are shown in Figure~\ref{fig:trends}. Figure~\ref{fig:algam} displays the alpha particle contribution to the TAE growth rate.  The growth rate is normalized to the on-axis Alfv\'{e}n frequency at 10 T such that growth rates can be compared between fields. Initially, the alpha particle contribution to the growth rate increases with magnetic field, which is in part a representation of increasing $\beta_\alpha$.  Then, the contribution flattens out or reduces near $\lambda = 1$, indicating a loss of a TAE resonance. Also note that at lower fields, the lower-$n$ modes have higher growth rates than their equivalent higher-$n$ modes, while at higher fields the modes are more comparable or the higher-$n$ modes are more unstable.

More direct evidence of the loss of the resonance can be found by normalizing the alpha particle growth rate to mode frequency at the given field and to $\beta_\alpha$, such that comparison to equation~\eqref{eq:alphagrow} is immediate. This quantity is plotted in Figure~\ref{fig:normalgam}. Somewhat below $\lambda = 1$, the trend in $\gamma_\alpha / \left( \omega \beta_\alpha \right)$ changes from whatever it was below to a notable decrease.  This is due to the loss of resonance, as will be further represented by the results in Section~\ref{sec:resonance}.  The decrease often begins below $\lambda = 1$ because of effects not included in the simplified passing-particle, localized mode analytic treatment; these effects allow particles close to, but not exactly, fulfilling the analytic resonance condition to exchange energy with the mode.

The contribution of deuterium to mode damping is shown in Figure~\ref{fig:dgam}; the trends in the tritium damping are identical. As expected, the damping does not show the strong $\lambda$-related changes that the alpha particle contribution does. 

Finally, the overall growth rate of the modes is plotted in Figure~\ref{fig:overallgam}. For all of the modes, the initial increase with $\lambda$ due to increasing $\beta_\alpha$ is visible. For the $n=21$ mode, the loss of resonance allows the mode to achieve stability after it was unstable. For the other modes, the loss of the resonance appears to substantially reduce the growth rate near $\lambda = 1.0$ relative to what it would be if the lower-field increase were still present, such that two modes are completely stable and the other is near marginal stability.  

While such alpha particle resonance loss cannot be experimentally tested in presently-existing devices, we note that the stabilization of NBI-driven AE activity due to changing effective temperature of beam ions and the resulting resonance loss has been recently experimentally observed on DIII-D \cite{pace2018dynamic}. Direct parallels can also be found in studies of the magnetic field dependence of  NBI-driven AEs on TFTR \cite{wong1991excitation}. While providing clear demonstration of resonance loss, these experimental demonstrations differ from alpha particle resonance loss in several fundamental ways, including the fast ion velocities at which they occur and the radial profiles and distribution of pitch angle characterizing the fast ions. 

\subsection{Energy of alpha particles interacting with mode}
\label{sec:resonance}
Next, we examine plots of regions in phase space of energy exchange between alpha particles and one of the TAEs produced by the CASTOR-K analysis.  This serves to demonstrate the energy of the alpha particles which interact with the modes, to give further evidence that the trends near $\lambda = 1.0$ noted in the previous section are indeed due to the loss of resonances, and to give insight into the different roles played by different types of particle orbits.

\begin{figure*}[!ht]
\centering
\subfigure[$\,$7 T, $\lambda = 0.74$, $\Lambda =0.01$]{\label{fig:passlow} \includegraphics[width=.9\columnwidth]{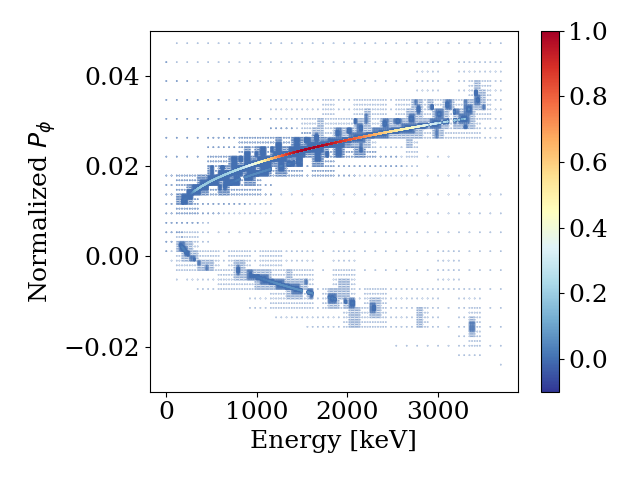}}
\subfigure[$\,$7 T, $\lambda = 0.74$, $\Lambda = 1.0$ ]{\label{fig:traplow}\includegraphics[width=.9\columnwidth]{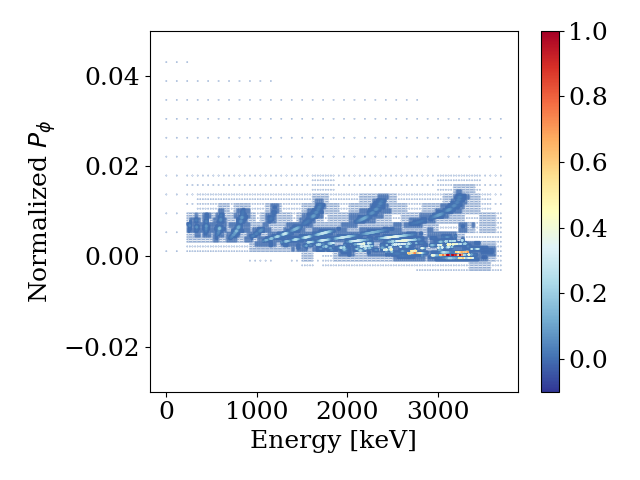}}
\subfigure[$\,$12 T, $\lambda = 0.97$, $\Lambda = 0.01$] {\label{fig:passmid}\includegraphics[width=.9\columnwidth]{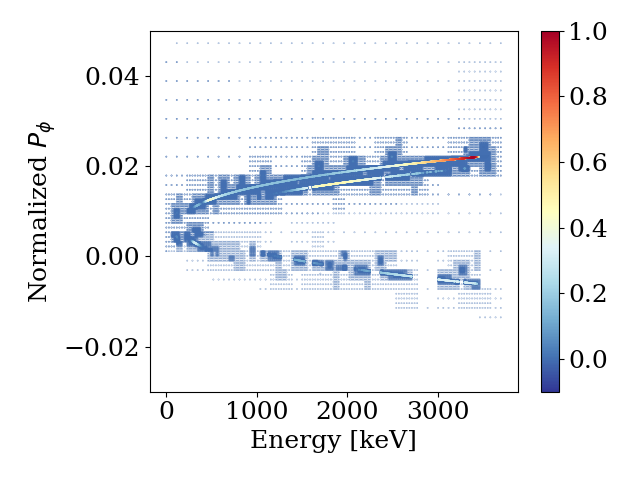}}
\subfigure[$\,$12 T, $\lambda = 0.97 $, $\Lambda = 1.0$ ] {\label{fig:trapmid}\includegraphics[width=.9\columnwidth]{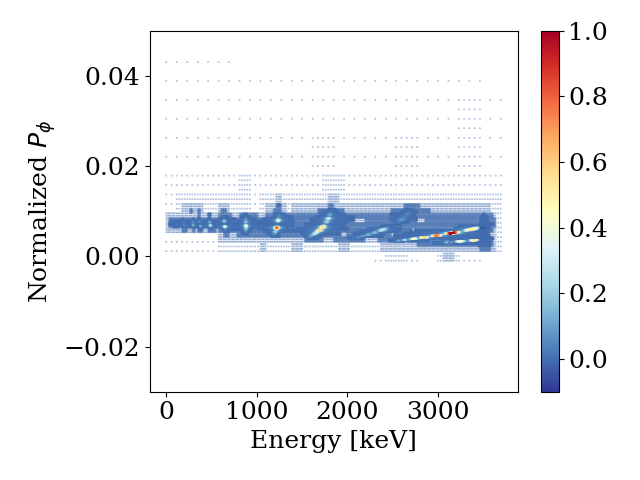}}
\subfigure[$\,$16 T, $\lambda = 1.11$, $\Lambda = 0.01$ ] {\label{fig:passhigh}\includegraphics[width=.9\columnwidth]{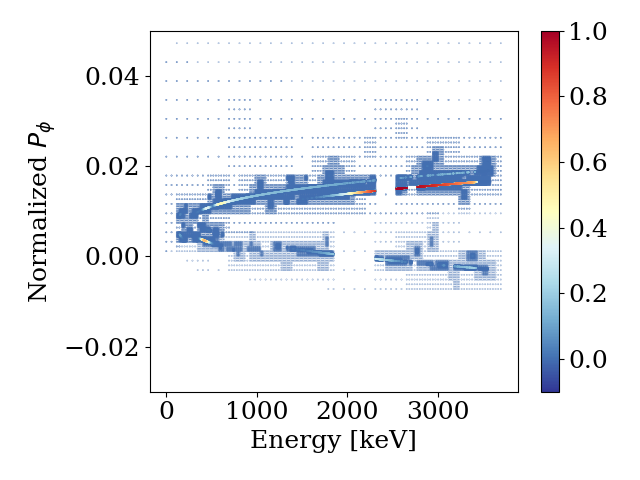}}
\subfigure[$\,$16 T, $\lambda = 1.11$, $\Lambda = 1.0$ ] {\label{fig:traphigh}\includegraphics[width=.9\columnwidth]{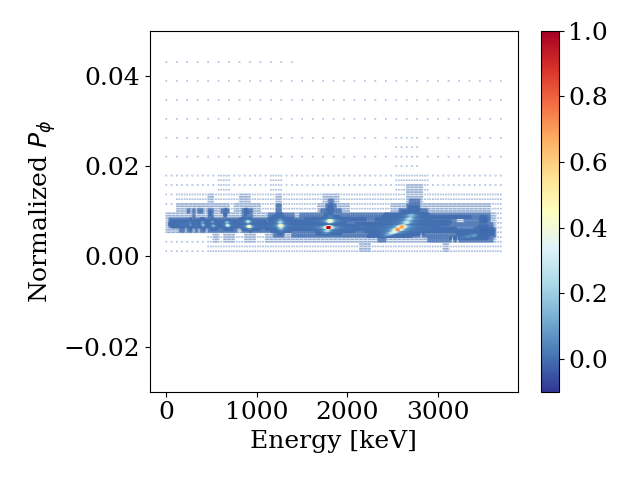}}
\caption{Regions of energy exchange between alpha particles and the TAE pictured in Figure~\ref{fig:thirdtae} at increasing magnetic fields and two values of $\Lambda$. Plots are normalized to the highest intensity of energy exchange at the particular field and value of $\Lambda$.}
\label{fig:resplot}
\end{figure*}

We consider results shown in Figure~\ref{fig:resplot}, which show energy exchange between the $n=22$ TAE pictured in Figure~\ref{fig:thirdtae} and the alpha particle population.  These results are shown at two different values of $\Lambda \equiv \left(v_\perp^2 / v^2 \right) \left( B_0 / B \right)$, $\Lambda = 0.01$ and $\Lambda = 1.0$.  The lower value of $\Lambda$ corresponds to passing particles, which are the only types of particles considered by the analytical theory presented in Section~\ref{sec:analytical}, while the higher value of $\Lambda$ corresponds to trapped particle orbits, and is usually one of the values of $\Lambda$ at which the most energy exchange between the mode and particles occurs.  The results are shown as a function of alpha particle energy and of toroidal canonical angular momentum $P_\phi \equiv q_\alpha \left( \psi  + v_\parallel R B_\phi / \Omega_\alpha \right)$  normalized to $q_\alpha B_0 R_0^2$.  The plots are evaluated at 7 T, 12 T, and 16 T, which correspond in this setup to $\lambda = 0.74, 0.97,$ and $1.11$.

Consider first the lowest-$\lambda$ plot for passing particles, Figure~\ref{fig:passlow}. This plot displays one primary resonance peaked roughly around $2000$ keV, corresponding to the primary TAE resonance $v_\parallel = v_{A0}$. This resonance is very spread out. The lower resonance, though present, is not significant. The corresponding plot for more complex particle orbit is shown in Figure~\ref{fig:traplow}; this plot has more complicated resonance structure. 

Next, consider the plots at 12 T, $\lambda = 0.97$, Figures~\ref{fig:passmid} and~\ref{fig:trapmid}. Notable in these plots is the partial cutoff of resonances near the alpha particle birth velocity, even in the case of more complex particle orbits. Lower energy resonances, including the $v_\parallel = v_{A0}/3$ resonance in Figure~\ref{fig:passmid} become noticeable as  a key component of the mode's drive, reducing the overall energy of the alpha particles that interact with the mode.

Finally, plots at 16 T, $\lambda = 1.11$ are shown in Figures~\ref{fig:passhigh} and~\ref{fig:traphigh}. In these plots, the primary resonance at $v_\parallel = v_{A0}$ for passing particles has nearly completely disappeared from the map. However, another resonance at high energy has appeared (which was slightly visible in~\ref{fig:passmid}), likely due to the complex, non-localized mode structure of the particular TAE examined, which has yet to be fully cut off. The lower energy resonances are very visible as a source of TAE drive. Likewise, in the $\Lambda = 1.0$ case, many resonances have been cut off and those remaining are at lower energy.

\subsection{Toroidal mode number of the most unstable mode}
\label{sec:tormodtrend}
\begin{figure*}[!ht]
\centering
\subfigure[$\,$7 T, $\lambda = 0.74$]{\label{} \includegraphics[width=.9\columnwidth]{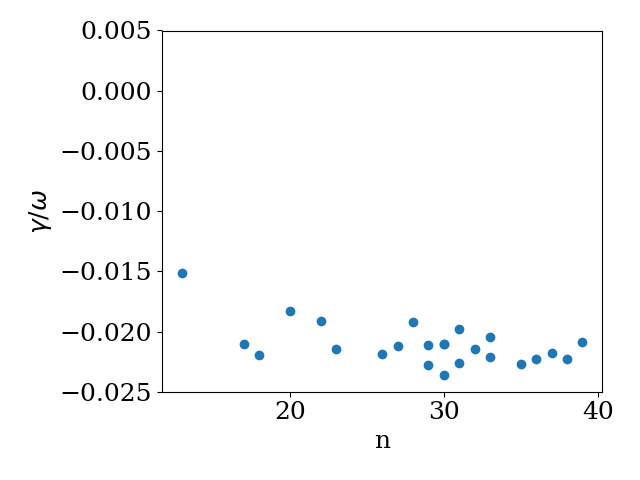}}
\subfigure[$\,$13 T, $\lambda = 1.00$ ]{\label{}\includegraphics[width=.9\columnwidth]{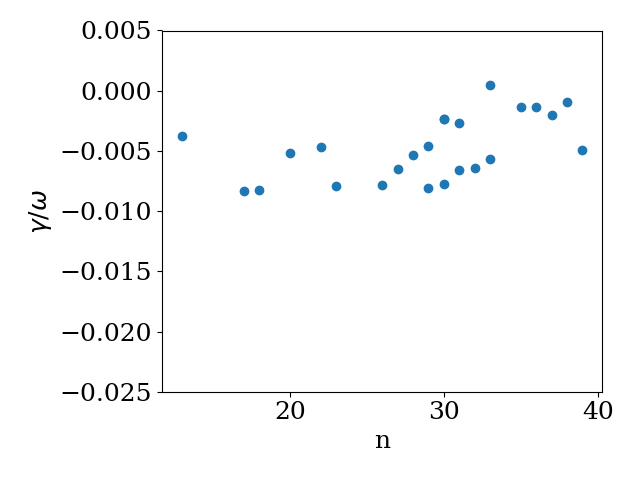}}
\caption{Mode growth rate normalized to mode frequency at two different magnetic fields, for TAEs with frequencies between $\omega = 0.45 \, \omega_{A0}$ and $\omega = 0.5 \, \omega_{A0}$.}
\label{fig:ntor}
\end{figure*}
Finally, we may consider trends in the toroidal mode number of the most unstable TAEs with magnetic field. Figure~\ref{fig:ntor} displays overall growth rate (including alpha particle, deuterium, and tritium contributions) as a function of toroidal mode number for modes with frequencies between $\omega = 0.45 \, \omega_{A0}$ and $\omega = 0.5 \, \omega_{A0}$ (because mode frequency is correlated to radial location, this allows a comparison between modes which experience similar background gradients and with similar structures).  At 7 T, the lower $n$ modes are closer to being unstable than the higher-$n$ modes (because $\beta_\alpha$ is low at this low field, none of the modes is unstable).  At 13 T, the higher-$n$ modes have higher overall growth rates (though few are unstable, in part because of the cut off of resonances at this field).

\subsection{Probable effect of finite alpha gyroradius on magnetic field trends}
\label{sec:probeff}
While the numerical results in this section offer significant insight into AE linear stability trends with magnetic field, they do not include finite alpha gyroradius effects.  These become important when $k_\perp \rho_\alpha \sim 1$, or, equivalently, as discussed in the introduction to this section, when $n\sim 45 \left(B_0 / 10 \, \rm T\right)$, making them important to some of the low-field results we present in this section. We comment in this section on the probable effect of these terms on the trends presented in this section, guided by the analytical estimate that finite gyroradius effects reduce alpha particle drive by a factor of $1/J_0^2\left(k_\perp \rho_\alpha\right)$, with $J_0$ the zeroth order Bessel function \cite{rodrigues2015systematic,gorelenkov1999fast}.

For a given mode, the importance of these corrections will tend to decrease with magnetic field, such that damping due to finite alpha gyroradius effects decreases, leading to an overall increase in growth rate with field. For alphas $\rho_\alpha = m_\alpha v_\perp / \left(q_\alpha B_0 \right)$, which for a given perpendicular velocity gives $\rho_\alpha \sim 1/B_0$. This decrease in gyroradius will result in decreasing $k_\perp \rho_\alpha$ through field, and hence a reduction in the factor $1/J_0^2\left(k_\perp \rho_\alpha\right)$ by which finite gyroradius effects reduce the growth rate. Since $k_\perp  \sim 1 / \Delta_m \sim n$ (consider Equation~\ref{eq:modew2}), this trend will be stronger for higher-$n$ modes, such that the trend identified in Section~\ref{sec:tormodtrend} is accentuated: stronger reduction in damping with field will further increase the instability of high-$n$ modes with respect to low-$n$ modes as field increases.

\section{Conclusion: Implications of findings for the high magnetic field approach and future work}
\label{sec:pow}
\subsection{Summary}
In this paper, we have considered AE stability behavior relevant to next generation very-high-magnetic-field tokamaks which are currently experiencing significant interest in fusion research. We began in Section~\ref{sec:deps} by considering the dominant trends with magnetic field of tokamak parameters relevant to AE stability, while noting that at a given magnetic field, a tokamak has flexibility to slightly modify these quantities. A result of this study was the realization in Section~\ref{sec:alphas} that $\beta_\alpha$ will tend to be significantly larger in high magnetic field devices.

Sections~\ref{sec:analytical} and~\ref{sec:computational} then argue theoretically  and computationally for trends in AE stability with magnetic field.  The increase in $\beta_\alpha$ with magnetic field will increase AE drive.  If an HTS device operates with sufficiently high Alfv\'{e}n speed, it may be able to cut off AE resonances, reducing mode drive and the energy of alpha particles interacting with the mode. Meanwhile, damping will not depend strongly on magnetic field. The structure of AEs will not tend to change with device magnetic field, and the toroidal mode number of the most strongly driven mode will increase.

\subsection{Implications of findings for the high magnetic field approach}

Given the strong increase in $\beta_{\alpha}$ with magnetic field, it is likely that HTS machines will experience AE activity, and planning for the management of this activity should be part of their design process. However, consideration of the source of the $\beta_\alpha$ growth with magnetic field gives context to how this challenge should be evaluated when evaluating the relative merits of low- and high-field approaches. When tokamak figures of merit are held constant as magnetic field is scaled, as we do in this paper, the fusion rate~\eqref{eq:S} increases with magnetic field strength, and this increasing fusion rate is responsible for the increasing $\beta_\alpha$.  If a fusion device aims to produce a certain amount of fusion power, small, high magnetic field machines will indeed have higher $\beta_\alpha$.  This is because larger, lower magnetic field machines will obtain the target amount of fusion power by operating with a smaller source rate~\eqref{eq:S}, and a lower $\beta_\alpha$, over a larger volume, while smaller, high magnetic field machines will operate with high source rate~\eqref{eq:S}, and a high $\beta_\alpha$ over a smaller volume. Since high $\beta_\alpha$  encourages AE instability, this is a critical consideration for high magnetic field machines.

However, if economic considerations are taken into account, this trend may not be overly punitive to high magnetic field devices.  Some studies \cite{fed} assume that any device, regardless of size, will for economic viability need to achieve not a fixed amount of power, but rather a fixed value of the fusion power density given in~\eqref{eq:highb}. To achieve this, a device must achieve a specific value of fusion source rate~\eqref{eq:S}, regardless of the device size and the resulting absolute amount of fusion power the device will produce. The tendency of AE modes to be destabilized is correlated to the ratio $\beta_\alpha/\beta$, which is roughly the ratio of AE drive to AE damping when the effect of resonance location is neglected (consider equations~\eqref{eq:thermaldamp} and~\eqref{eq:alphagrow}). A figure of merit may be defined which represents the amount of $\beta_\alpha/ \beta$ per amount of fusion power density:
\begin{equation}
\label{eq:fixpov}
\frac{\beta_\alpha/ \beta}{n_D n_T \left< \sigma v \right>_{DT} } \sim \frac{\frac{n^2 T^2/\left(n/T^{3/2}\right)}{nT}}{n^2T^2} \sim  \frac{T^{1/2}}{n^2}.
\end{equation}
(Here, we have taken the simplified expression $\left<\sigma v \right>_{DT} \sim T^2$ and neglected the influence of alpha pressure on $\beta$.)
This quantity strongly decreases with $B_0$ due to the higher absolute densities accessible at high magnetic field.  Thus, when economic considerations are taken into account, the amount of AE drive incurred to achieve a given performance decreases with field. This is a positive trend for high magnetic field machines.

Regardless of the lens through which the high field approach is evaluated, the results of this paper suggest two 
regimes in which the unique characteristics of HTS machines are optimally exploited  to minimize and control AE activity that is present: at low density and high density. Operation in the first of these regimes attempts to cut off AE resonances, while the latter accelerates the slowing down of alphas. Intermediate densities allow neither of these advantages. Which of these regimes an HTS machine operates in is likely to be determined by  operational constraints distinct from AE behavior. We consider each of these regimes in turn here.

Cutting off AE resonances, even partially, reduces the drive to AEs and the energy of alpha particles which interact with the mode. This will reduce the ability of the AEs to cause  transport, though significant transport will likely still be present.  However, cutting off even the first TAE resonance, i.e. achieving $\lambda \sim 1$, at practically-achievable magnetic fields, will require operating at low Greenwald fraction and potentially high values of $q_{95}$ (and thus lower values of current). Consider that the configuration studied in Section~\ref{sec:computational}, with $q_{95} = 3.5$ and $f_{GW} = 0.18$, only achieves $\lambda = 1$ at 13 T.  
We note that such low Greenwald fraction at high field corresponds to significant density; in the scenario considered, the density at 10 T (see Figure~\ref{fig:profs}) is higher than that in the ITER baseline scenario \cite{rodrigues2015systematic}. Correspondingly, such low Greenwald fractions are within the operating space currently envisioned for high field devices\cite{psfcreport}.
 However, because of favorable trends in overall tokamak performance with increasing current and density, such low Greenwald fraction operation may not be the preferred operation state for high-magnetic-field devices, and certainly does not represent the highest performance they could ideally achieve.  Nevertheless, operation in such a low density state may end up being a requirement of high-magnetic-field operation if the trends in H-mode density pedestal height, mentioned in Section~\ref{sec:deps}, or perhaps a similar trend in H-mode density limit, are born out at high fields, or if I-mode operation is required due to some operational constraint.  Even if not practical in high-performance shots, resonance cutoff could still be an interesting physics effect to observe; furthermore, given the broad nature of the resonances and the resulting beginning of resonance cutoff well before $\lambda = 1$ for some modes, it will likely occur for a few modes even in higher Greenwald-fraction shots. 

If a device does not operate at low density due to operational or mission constraints, it can minimize AE drive by maximizing density at the expense of temperature. A given device will want to minimize AE drive subject to a fixed performance.  Neglecting the effect of resonance locations, as in~\eqref{eq:fixpov}, this means minimizing $\beta_\alpha/\beta \sim T^{5/2}$ subject to $n^2T^2 = C$, with $C$ a constant. This corresponds to minimizing
\begin{equation}
\beta_\alpha \sim \frac{C^{5/4}}{n^{5/2}}.
\end{equation}
This division by density exists because higher-density plasmas slow down alpha particles more quickly, and indicates that increasing $n$ while decreasing $T$ at fixed $\beta$ will reduce the amount of AE drive.  This trend was also noted in Section~\ref{sec:alphas}. High field machines are better able to make this tradeoff than low field machines, because high field machines can access higher densities due to their higher currents, when compared to low-field devices with similar power densities.

\subsection{Future work}

The present work suggests several avenues for future work. First, as noted in Section~\ref{sec:heating}, further analysis of heating drive for AEs should be completed. We observe that this may be particularly important for the SPARC concept, as it has a fairly modest $Q$ target of 2-5 \cite{psfcreport},  meaning that a significant fraction of AE drive could come from fast ions resulting from ICRH or other heating methods.

Next, this paper has focused on the TAE, and its results may be applied in a straightforward way to the EAE, NAE, and other gap modes. However, a variety of other Alfv\'{e}nic fluctuations can be excited in tokamaks. They are not a focus of this paper due to their different physical nature, which  prevents them from being studied using the suite of codes employed here, which is not capable of treating continuum damping or thermal plasma compressibility. In addition, some modes, like the RSAE (caused by a reversed shear $q$ profile), are left out because they are not expected to be observed in baseline scenarios for SPARC, which are currently envisioned to use conservative profiles without reversed shear \cite{greenwald2018performance}.

However, study of these modes and the dependence of their behavior on magnetic field strength is important and should be undertaken in detail in future work. In many cases, the framework established in this paper can be  adapted to other modes; we give two examples here. Energetic particle continuum modes (EPMs) are modes at frequencies characteristic of energetic particle motion (often outside of the gaps in the continuum spectrum, where the TAE and similar modes live), which arise when drive from the spatial gradient in the fast particles is sufficient to exceed continuum damping \cite{0029-5515-47-6-S05}. The excitation of these modes is expected to be only weakly dependent on the velocity space fast particle distribution function \cite{0029-5515-47-6-S05}. The smaller amount of fast ion beta generated at a given fusion power density thus is likely to be the most important determinant of EPM stability. This will prove advantageous for high-magnetic field reactors, and indeed previous studies of high-field devices like IGNITOR and FIRE have found these devices to be stable to EPMs \cite{vlad2004consistency}. 

Beta-induced Alfv\'{e}n eigenmodes (BAEs) exist in a frequency gap caused by finite thermal plasma compressibility located below the shear Alfv\'{e}n continuum spectrum,  $0 < \left(\omega/\omega_{A0}\right)^2 \lesssim \Gamma \beta$, with $\Gamma$ the ratio of specific heats \cite{zonca1996kinetic}. Analytically, these modes can be shown to be resonant with energetic particles with transit frequencies approximately equal to  the mode frequency; the source of the drive is the spatial gradient of the energetic particle distribution function \cite{zonca1996kinetic,ma2014linear}.  That the resonant velocities for these modes are significantly lower than for the TAE suggests it will not be possible to cut them off, and for this mode, again, the magnetic field dependence of $\beta_\alpha$ is the most important determinant of alpha particle contribution to the growth rate.

Finally, the topic of this paper is linear AE behavior at high field, but actual alpha particle transport and evolution results from nonlinear processes \cite{heidbrink2008basic}. Comparable analysis of high field trends in this area is a topic of current research by the authors. In addition to trends intrinsic to nonlinear physics, this analysis could reveal nonlinear trends that result from linear physics.  For example, high $n$ modes, which tend to be narrower, are more likely than low $n$ modes to be unstable in high field machines, while the converse is true in low field machines. The reduced width of destabilized AEs in HTS machines could influence the ability of these modes to cause radial transport; further, the interaction of AEs with periodic magnetic field ripple is a key source alpha transport \cite{heidbrink2008basic,pinches2015energetic}, and the changed periodicity of destabilized AEs could influence this synergy.

\section{Acknowledgments}
The authors acknowledge support from the National Science Foundation Graduate Research Fellowship under Grant No. DGE-1122374, US Department of Energy awards \uppercase{DE-SC}0014264 and DE-FG02-91ER54109, and Funda\c{c}\~{a}o para a Ci\^{e}ncia e a Tecnologia (FCT, Lisbon) project UID/FIS/50010/2013. This research used resources of the National Energy Research Scientific Computing Center (NERSC), a U.S. Department of Energy Office of Science User Facility operated under Contract No. DE-AC02-05CH11231.  The authors thank Syun'ichi Shiraiwa, Yijun Lin, Theodore Golfinopoulos, Lucio Milanese, and Steven Wukitch for helpful conversations. In addition, the authors thank D. Borba of the Theory and Modelling group of Instituto de Plasmas e Fus\~{a}o Nuclear for providing the code CASTOR-K, and S. Sharapov of Culham Centre for Fusion Energy for providing the codes HELENA and MISHKA.

\appendix
\section{Slowing down distribution and modification for two plasma species}
\label{sec:slowing}
We model fusion alphas using a slowing down distribution modified to account for two bulk species. In this section, we outline the derivation of the one-species slowing down distribution (details of which may be found in Ref. \cite{goldston1995introduction}), then modify it for two bulk species.  

The distribution is derived from a Fokker-Planck equation for slowing down due to collisions with electrons and ions,
\begin{equation}
\left(\frac{\partial f_b}{\partial t } \right)_{col}  = - \frac{\partial}{\partial \vec{v}} \cdot \left( \frac{d  \left< \Delta\vec{v} \right>}{dt} f_b \right),
\end{equation}
with $f_b\left(\vec{v},t\right)$ the distribution of particles in a beam; the quantity $d \left< \Delta \vec{v} \right> / dt $ represents the average rate of change due to Coulomb collisions of a particle's mean directed velocity.                                                    
The right hand side of the equation may be derived for components due to both collisions with ions and electrons, giving
\begin{equation}
\frac{\partial f_b}{\partial t} = \frac{n_e Z Z_b^2 e^4 \ln \Lambda}{4 \pi \epsilon_0^2 M_b M } \frac{\partial}{\partial \vec{v}} \cdot  \left[ \frac{\vec{v}}{v^3} \left( 1 + \frac{v^3}{v_{crit}^3}\right) f_b \right],
\end{equation}
where 
\begin{equation}
v_{crit} = 3^{1/3} Z^{1/3} \left( \pi /2 \right)^{1/6} \left[ T_e / \left(m_e^{1/3} M^{2/3} \right)\right]^{1/2},
\end{equation}
with $M_b$ the beam ion mass, $m_e$  the electron mass and $M$ the bulk ion mass.

To describe alpha particles, one transforms to spherical coordinates because alpha birth is isotropic:
\begin{equation}
\frac{\partial f_b }{\partial t} = \frac{n_e Z Z_b^2 e^4 \ln \Lambda}{4 \pi \epsilon_0^2 M_b M} \frac{1}{v^2} \frac{\partial}{\partial v} \left[ \left( 1 + \frac{v^3}{v_{crit}^3}\right) f_b \right].
\end{equation}
Now, consider that alphas are ``injected" into the plasma with birth velocity $v_{\alpha 0}$.  This adds a source 
\begin{equation}
\left( \frac{\partial f_b}{ \partial t} \right)_{\textrm{source}} = \frac{S \delta \left(v-v_{\alpha 0} \right)}{4 \pi v^2}
\end{equation}
with $S$ the alpha birth rate (note that the normalization is chosen such that integration over all velocity space with volume element $4 \pi v^2 dv$ yields growth at the source rate). So, for steady state, slowing down alphas will be described by
\begin{equation}
0 =  \frac{S \delta \left(v-v_{\alpha 0} \right)}{4 \pi v^2} + \frac{n_e Z Z_b^2 e^4 \ln \Lambda}{4 \pi \epsilon_0^2 M_b M} \frac{1}{v^2} \frac{\partial}{\partial v} \left[ \left( 1 + \frac{v^3}{v_{crit}^3}\right) f_b \right].
\end{equation}
Solving this equation gives 
\begin{equation}
\label{eq:dist1}
f_b = \frac{S \epsilon_0^2 M M_b}{n_e Z Z_b^2 e^4 \ln \Lambda} \left( \frac{1}{1 + v^3/v_{crit}^3}\right), \, v<v_{\alpha 0}.
\end{equation}
Note that in this formulation the particles are lost in an effect sink at $v = 0$ which corresponds to them transitioning to Maxwellian helium ash.

This derivation is for slowing down on one species of ion.  In this paper, the plasma is primarily composed of two bulk ions, deuterium and tritium.  The previous derivation can be straightforwardly modified for two species. Consider that the derivation relies on the statement of the dynamical friction due to ions of 
\begin{equation}
\label{eq:dynfic1}
\frac{d \left<\vec{v} \right>}{dt}= \frac{-n_i Z^2 Z_b^2 e^4 \ln \Lambda}{4 \pi \epsilon_0^2 M M_b v^3}\vec{v}.
\end{equation}
With two ion species of masses $M_1$ and $M_2$ and densities $n_{i,1}$ and $n_{i,2}$, two contributions of this sort are combined to read
\begin{equation}
\frac{d \left< \vec{v} \right>}{dt}= \frac{- Z^2 Z_b^2 e^4 \ln \Lambda}{4 \pi \epsilon_0^2 M_b v^3}\vec{v} \left(\frac{n_{i,1}}{M_1} +\frac{n_{i,2}}{M_2} \right).
\end{equation}
With $n_{i,1} = n_{i,2} = n_i/2$, this  becomes 
\begin{equation}
\label{eq:dynfic2}
\frac{d \left<\vec{v} \right>}{dt}= \frac{- Z^2 Z_b^2 e^4 \ln \Lambda}{4 \pi \epsilon_0^2 M_b v^3}\vec{v} \left( \frac{n_i}{2 \left( \frac{M_1 M_2}{M_1+M_2}\right)}\right).
\end{equation}
Comparing~\eqref{eq:dynfic1} and~\eqref{eq:dynfic2}, we see that the one-species slowing down distribution can be modified for two species by making the replacement
\begin{equation}
M \rightarrow \frac{2M_1 M_2}{M_1 + M_2} \equiv m_{equiv}.
\end{equation}

\bibliographystyle{unsrtnat}
\bibliography{cmodbib}

\begin{thebibliography}{57}
\providecommand{\natexlab}[1]{#1}
\providecommand{\url}[1]{\texttt{#1}}
\expandafter\ifx\csname urlstyle\endcsname\relax
  \providecommand{\doi}[1]{doi: #1}\else
  \providecommand{\doi}{doi: \begingroup \urlstyle{rm}\Url}\fi

\bibitem[Wesson and Campbell(2011)]{wesson2004tokamaks}
J.~Wesson and D.J. Campbell.
\newblock \emph{Tokamaks}.
\newblock Oxford University Press, 2011.

\bibitem[Sorbom et~al.(2015)Sorbom, Ball, Palmer, Mangiarotti, Sierchio,
  Bonoli, Kasten, Sutherland, Barnard, Haakonsen, et~al.]{sorbom2015arc}
B.N. Sorbom, J.~Ball, T.R. Palmer, F.J. Mangiarotti, J.M. Sierchio, P.~Bonoli,
  C.~Kasten, D.A. Sutherland, H.S. Barnard, C.B. Haakonsen, et~al.
\newblock {ARC}: A compact, high-field, fusion nuclear science facility and
  demonstration power plant with demountable magnets.
\newblock \emph{Fusion Engineering and Design}, 100:\penalty0 378--405, 2015.

\bibitem[Hughes et~al.(2018)Hughes, Snyder, Reinke, LaBombard, Mordijck, Scott,
  Tolman, Baek, Golfinopoulos, Granetz, et~al.]{hughes2018access}
J.W. Hughes, P.B. Snyder, M.L. Reinke, B.~LaBombard, S.~Mordijck, S.~Scott,
  E.A. Tolman, S.G. Baek, T.~Golfinopoulos, R.~Granetz, et~al.
\newblock Access to pedestal pressure relevant to burning plasmas on the high
  magnetic field tokamak {A}lcator {C}-{M}od.
\newblock \emph{Nuclear Fusion}, 2018.
\newblock URL \url{https://doi.org/10.1088/1741-4326/aabc8a}.

\bibitem[Troyon et~al.(1984)Troyon, Gruber, Saurenmann, Semenzato, and
  Succi]{troyon1984mhd}
F.~Troyon, R.~Gruber, H.~Saurenmann, S.~Semenzato, and S.~Succi.
\newblock {MHD}-limits to plasma confinement.
\newblock \emph{Plasma Physics and Controlled Fusion}, 26\penalty0
  (1A):\penalty0 209, 1984.

\bibitem[Fietz et~al.(2005)Fietz, Fink, Heller, Komarek, Tanna, Zahn, Pasztor,
  Wesche, Salpietro, and Vostner]{fietz2005high}
W.H. Fietz, S.~Fink, R.~Heller, P.~Komarek, V.L. Tanna, G.~Zahn, G.~Pasztor,
  R.~Wesche, E.~Salpietro, and A.~Vostner.
\newblock High temperature superconductors for the {ITER} magnet system and
  beyond.
\newblock \emph{Fusion Engineering and Design}, 75:\penalty0 105--109, 2005.

\bibitem[Greenwald et~al.(2018{\natexlab{a}})Greenwald, Whyte, Bonoli, Hartwig,
  Irby, LaBombard, Marmar, Minervini, Takayasu, Terry, Vieira, White, Wukitch,
  Brunner, Mumgaard, and Sorbom]{psfcreport}
M.~Greenwald, D.~Whyte, P.~Bonoli, Z.~Hartwig, J.~Irby, B.~LaBombard,
  E.~Marmar, J.~Minervini, M.~Takayasu, J.~Terry, R.~Vieira, A.~White,
  S.~Wukitch, D.~Brunner, R.~Mumgaard, and B.~Sorbom.
\newblock The high-field path to practical fusion energy.
\newblock \emph{PSFC Report}, RR-18-2, 2018{\natexlab{a}}.

\bibitem[Shimada et~al.(2007)Shimada, Campbell, Mukhovatov, Fujiwara, Kirneva,
  Lackner, Nagami, Pustovitov, Uckan, Wesley, et~al.]{shimada2007overview}
M.~Shimada, D.J. Campbell, V.~Mukhovatov, M.~Fujiwara, N.~Kirneva, K.~Lackner,
  M.~Nagami, V.D. Pustovitov, N.~Uckan, J.~Wesley, et~al.
\newblock Progress in the {ITER} {P}hysics {B}asis {C}hapter 1: Overview and
  summary.
\newblock \emph{Nuclear Fusion}, 47\penalty0 (6):\penalty0 S1, 2007.

\bibitem[Heidbrink(2008)]{heidbrink2008basic}
W.W. Heidbrink.
\newblock Basic physics of {A}lfv{\'e}n instabilities driven by energetic
  particles in toroidally confined plasmas).
\newblock \emph{Physics of Plasmas}, 15\penalty0 (5):\penalty0 055501, 2008.

\bibitem[Gorelenkov et~al.(2003)Gorelenkov, Berk, Budny, Cheng, Fu, Heidbrink,
  Kramer, Meade, and Nazikian]{gorelenkov2003study}
N.N. Gorelenkov, H.L. Berk, R.~Budny, C.Z. Cheng, G-Y. Fu, W.W. Heidbrink, G.J.
  Kramer, D.~Meade, and R.~Nazikian.
\newblock Study of thermonuclear {A}lfv{\'e}n instabilities in next step
  burning plasma proposals.
\newblock \emph{Nuclear Fusion}, 43\penalty0 (7):\penalty0 594, 2003.

\bibitem[Rodrigues et~al.(2016)Rodrigues, Figueiredo, Borba, Coelho,
  Fazendeiro, Ferreira, Loureiro, Nabais, Pinches, Polevoi,
  et~al.]{rodrigues2016sensitivity}
P.~Rodrigues, A.C.A. Figueiredo, D.~Borba, R.~Coelho, L.~Fazendeiro,
  J.~Ferreira, N.F. Loureiro, F.~Nabais, S.D. Pinches, A.R. Polevoi, et~al.
\newblock Sensitivity of alpha-particle-driven {A}lfv{\'e}n eigenmodes to
  q-profile variation in {ITER} scenarios.
\newblock \emph{Nuclear Fusion}, 56\penalty0 (11):\penalty0 112006, 2016.

\bibitem[Yang et~al.(2017)Yang, Li, Hu, and Gao]{yang2017linear}
W.~Yang, G.~Li, Y.~Hu, and X.~Gao.
\newblock Linear stability of toroidal {A}lfv{\'e}n eigenmodes in the {C}hinese
  {F}usion {E}ngineering {T}est {R}eactor.
\newblock \emph{Fusion Engineering and Design}, 114:\penalty0 118--126, 2017.

\bibitem[Chen et~al.(2010)Chen, Parker, Lang, and Fu]{chen2010linear}
Y.~Chen, S.E. Parker, J.~Lang, and G.Y. Fu.
\newblock Linear gyrokinetic simulation of high-n toroidal {A}lfv{\'e}n
  eigenmodes in a burning plasma.
\newblock \emph{Physics of Plasmas}, 17\penalty0 (10):\penalty0 102504, 2010.

\bibitem[Rodrigues et~al.(2015)Rodrigues, Figueiredo, Ferreira, Coelho, Nabais,
  Borba, Loureiro, Oliver, and Sharapov]{rodrigues2015systematic}
P.~Rodrigues, A.~Figueiredo, J.~Ferreira, R.~Coelho, F.~Nabais, D.~Borba, N.F.
  Loureiro, H.J.C. Oliver, and S.E. Sharapov.
\newblock Systematic linear-stability assessment of {A}lfv{\'e}n eigenmodes in
  the presence of fusion $\alpha$-particles for {ITER}-like equilibria.
\newblock \emph{Nuclear Fusion}, 55\penalty0 (8):\penalty0 083003, 2015.

\bibitem[Figueiredo et~al.(2016)Figueiredo, Rodrigues, Borba, Coelho,
  Fazendeiro, Ferreira, Loureiro, Nabais, Pinches, Polevoi,
  et~al.]{figueiredo2016comprehensive}
A.~Figueiredo, P.~Rodrigues, D.~Borba, R.~Coelho, L.~Fazendeiro, J.~Ferreira,
  N.F. Loureiro, F.~Nabais, S.D. Pinches, A.R. Polevoi, et~al.
\newblock Comprehensive evaluation of the linear stability of {A}lfv{\'e}n
  eigenmodes driven by alpha particles in an {ITER} baseline scenario.
\newblock \emph{Nuclear Fusion}, 56\penalty0 (7):\penalty0 076007, 2016.

\bibitem[Jaun et~al.(2000)Jaun, Fasoli, Vaclavik, and
  Villard]{jaun2000stability}
A.~Jaun, A.~Fasoli, J.~Vaclavik, and L.~Villard.
\newblock Stability of {A}lfv{\'e}n eigenmodes in optimized tokamaks.
\newblock \emph{Nuclear Fusion}, 40\penalty0 (7):\penalty0 1343, 2000.

\bibitem[Grad and Rubin(1958)]{grad1958hydromagnetic}
H.~Grad and H.~Rubin.
\newblock Hydromagnetic equilibria and force-free fields.
\newblock \emph{Journal of Nuclear Energy (1954)}, 7\penalty0 (3-4):\penalty0
  284--285, 1958.

\bibitem[Shafranov(1960)]{shafranov1960equilibrium}
V.D. Shafranov.
\newblock Equilibrium of a plasma toroid in a magnetic field.
\newblock \emph{Soviet Physics JETP-USSR}, 10\penalty0 (4):\penalty0 775--779,
  1960.

\bibitem[L{\"u}st and Schl{\"u}ter(1957)]{lust1957axialsymmetrische}
R~L{\"u}st and A~Schl{\"u}ter.
\newblock Axialsymmetrische magnetohydrodynamische
  gleichgewichtskonfigurationen.
\newblock \emph{Zeitschrift f{\"u}r Naturforschung A}, 12\penalty0
  (10):\penalty0 850--854, 1957.

\bibitem[Neu et~al.(2007)Neu, Balden, Bobkov, Dux, Gruber, Herrmann,
  Kallenbach, Kaufmann, Maggi, Maier, et~al.]{neu2007plasma}
R.~Neu, M.~Balden, V.~Bobkov, R.~Dux, O.~Gruber, A.~Herrmann, A.~Kallenbach,
  M.~Kaufmann, C.F. Maggi, H.~Maier, et~al.
\newblock Plasma wall interaction and its implication in an all tungsten
  divertor tokamak.
\newblock \emph{Plasma Physics and Controlled Fusion}, 49\penalty0
  (12B):\penalty0 B59, 2007.

\bibitem[Greenwald et~al.(1988)Greenwald, Terry, Wolfe, Ejima, Bell, Kaye, and
  Neilson]{greenwald1988new}
M.~Greenwald, J.L. Terry, S.M. Wolfe, S.~Ejima, M.G. Bell, S.M. Kaye, and G.H.
  Neilson.
\newblock A new look at density limits in tokamaks.
\newblock \emph{Nuclear Fusion}, 28\penalty0 (12):\penalty0 2199, 1988.

\bibitem[Hughes et~al.(2002)Hughes, Mossessian, Hubbard, LaBombard, and
  Marmar]{hughes2002observations}
J.W. Hughes, D.A. Mossessian, A.E. Hubbard, B.~LaBombard, and E.S. Marmar.
\newblock Observations and empirical scalings of the high-confinement mode
  pedestal on {A}lcator {C}-{M}od.
\newblock \emph{Physics of Plasmas}, 9\penalty0 (7):\penalty0 3019--3030, 2002.

\bibitem[Hughes et~al.(2013)Hughes, Snyder, Walk, Davis, Diallo, LaBombard,
  Baek, Churchill, Greenwald, Groebner, et~al.]{hughes2013pedestal}
J.W. Hughes, P.B. Snyder, J.R. Walk, E.M. Davis, A.~Diallo, B.~LaBombard, S.G.
  Baek, R.M. Churchill, M.~Greenwald, R.J. Groebner, et~al.
\newblock Pedestal structure and stability in {H}-mode and {I}-mode: a
  comparative study on {A}lcator {C}-{M}od.
\newblock \emph{Nuclear Fusion}, 53\penalty0 (4):\penalty0 043016, 2013.

\bibitem[Hughes et~al.(2007)Hughes, LaBombard, Terry, Hubbard, and
  Lipschultz]{hughes2007edge}
J.W. Hughes, B.~LaBombard, J.~Terry, A.E. Hubbard, and B.~Lipschultz.
\newblock Edge profile stiffness and insensitivity of the density pedestal to
  neutral fueling in {A}lcator {C}-{M}od edge transport barriers.
\newblock \emph{Nuclear Fusion}, 47\penalty0 (8):\penalty0 1057, 2007.

\bibitem[Tolman et~al.(2018)Tolman, Hughes, Wolfe, Wukitch, LaBombard, Hubbard,
  Marmar, Snyder, and Schmidtmayr]{tolman2018influence}
E.A. Tolman, J.W. Hughes, S.M. Wolfe, S.J. Wukitch, B.~LaBombard, A.E. Hubbard,
  E.S. Marmar, P.B. Snyder, and M.~Schmidtmayr.
\newblock Influence of high magnetic field on access to stationary {H}-modes
  and pedestal characteristics in {A}lcator {C}-{M}od.
\newblock \emph{Nuclear Fusion}, 58\penalty0 (4):\penalty0 046004, 2018.

\bibitem[Greenwald(2002)]{greenwald2002density}
M.~Greenwald.
\newblock Density limits in toroidal plasmas.
\newblock \emph{Plasma Physics and Controlled Fusion}, 44\penalty0
  (8):\penalty0 R27, 2002.

\bibitem[Borrass et~al.(2004)Borrass, Loarte, Maggi, Mertens, Monier, Monk,
  Ongena, Rapp, Saibene, Sartori, et~al.]{borrass2004recent}
K.~Borrass, A.~Loarte, C.F. Maggi, V.~Mertens, P.~Monier, R.~Monk, J.~Ongena,
  J.~Rapp, G.~Saibene, R.~Sartori, et~al.
\newblock Recent {H}-mode density limit studies at {JET}.
\newblock \emph{Nuclear Fusion}, 44\penalty0 (7):\penalty0 752, 2004.

\bibitem[Bernert et~al.(2014)Bernert, Eich, Kallenbach, Carralero, Huber, Lang,
  Potzel, Reimold, Schweinzer, Viezzer, et~al.]{bernert2014h}
M.~Bernert, T.~Eich, A.~Kallenbach, D.~Carralero, A.~Huber, P.T. Lang,
  S.~Potzel, F.~Reimold, J.~Schweinzer, E.~Viezzer, et~al.
\newblock The {H}-mode density limit in the full tungsten {ASDEX} {U}pgrade
  tokamak.
\newblock \emph{Plasma Physics and Controlled Fusion}, 57\penalty0
  (1):\penalty0 014038, 2014.

\bibitem[Betti and Freidberg(1992)]{betti1992stability}
R.~Betti and J.P. Freidberg.
\newblock Stability of {A}lfv{\'e}n gap modes in burning plasmas.
\newblock \emph{Physics of Fluids B: Plasma Physics}, 4\penalty0 (6):\penalty0
  1465--1474, 1992.

\bibitem[Breizman and Sharapov(1995)]{breizman1995energetic}
B.N. Breizman and S.E. Sharapov.
\newblock Energetic particle drive for toroidicity-induced {A}lfv{\'e}n
  eigenmodes and kinetic toroidicity-induced {A}lfv{\'e}n eigenmodes in a
  low-shear tokamak.
\newblock \emph{Plasma Physics and Controlled Fusion}, 37\penalty0
  (10):\penalty0 1057, 1995.

\bibitem[Rosenbluth et~al.(1992)Rosenbluth, Berk, Van~Dam, and
  Lindberg]{rosenbluth1992continuum}
M.N. Rosenbluth, H.L. Berk, J.W. Van~Dam, and D.M. Lindberg.
\newblock Continuum damping of high-mode-number toroidal {A}lfv{\'e}n waves.
\newblock \emph{Physical Review Letters}, 68\penalty0 (5):\penalty0 596, 1992.

\bibitem[Pinches(1996)]{pinches1996nonlinear}
S.D. Pinches.
\newblock \emph{Nonlinear interaction of fast particles with {A}lfv{\'e}n waves
  in tokamaks}.
\newblock PhD thesis, University of Nottingham, 1996.

\bibitem[Heidbrink et~al.(2003)Heidbrink, Fredrickson, Gorelenkov, Hyatt,
  Kramer, and Luo]{heidbrink2003alfven}
W.W. Heidbrink, E.~Fredrickson, N.N. Gorelenkov, A.W. Hyatt, G.~Kramer, and
  Y.~Luo.
\newblock An {A}lfv{\'e}n eigenmode similarity experiment.
\newblock \emph{Plasma Physics and Controlled Fusion}, 45\penalty0
  (6):\penalty0 983, 2003.

\bibitem[Berk et~al.(1992)Berk, Breizman, and Ye]{berk1992finite}
H.L. Berk, B.N. Breizman, and H.~Ye.
\newblock Finite orbit energetic particle linear response to toroidal
  {A}lfv{\'e}n eigenmodes.
\newblock \emph{Physics Letters A}, 162\penalty0 (6):\penalty0 475--481, 1992.

\bibitem[Fu and Cheng(1992)]{fu1992excitation}
G.Y. Fu and C.Z. Cheng.
\newblock Excitation of high-n toroidicity-induced shear {A}lfv{\'e}n
  eigenmodes by energetic particles and fusion alpha particles in tokamaks.
\newblock \emph{Physics of Fluids B: Plasma Physics}, 4\penalty0 (11):\penalty0
  3722--3734, 1992.

\bibitem[Heidbrink(2002)]{heidbrink2002alpha}
W.W. Heidbrink.
\newblock Alpha particle physics in a tokamak burning plasma experiment.
\newblock \emph{Physics of Plasmas}, 9\penalty0 (5):\penalty0 2113--2119, 2002.

\bibitem[Martin et~al.(2008)Martin, Takizuka, et~al.]{martin2008power}
Y.R. Martin, T.~Takizuka, et~al.
\newblock Power requirement for accessing the {H}-mode in {ITER}.
\newblock In \emph{Journal of Physics: Conference Series}, volume 123, page
  012033. IOP Publishing, 2008.

\bibitem[Schmidtmayr et~al.(2018)Schmidtmayr, Hughes, Ryter, Wolfrum, Cao,
  Creely, Howard, Hubbard, Lin, Reinke, et~al.]{schmidtmayr2018investigation}
M.~Schmidtmayr, J.W. Hughes, F.~Ryter, E.~Wolfrum, N.~Cao, A.J. Creely,
  N.~Howard, A.E. Hubbard, Y.~Lin, M.L. Reinke, et~al.
\newblock Investigation of the critical edge ion heat flux for {L-H}
  transitions in {A}lcator {C}-{M}od and its dependence on {$B_T$}.
\newblock \emph{Nuclear Fusion}, 58\penalty0 (5):\penalty0 056003, 2018.

\bibitem[Mantsinen et~al.(2000)Mantsinen, Sharapov, Alper, Gondhalekar, and
  McDonald]{mantsinen2000new}
M.J. Mantsinen, S.~Sharapov, B.~Alper, A.~Gondhalekar, and D.C. McDonald.
\newblock A new type of {MHD} activity in jet {ICRF}-only discharges with high
  fast-ion energy contents.
\newblock \emph{Plasma Physics and Controlled Fusion}, 42\penalty0
  (12):\penalty0 1291, 2000.

\bibitem[Huysmans et~al.(1991)Huysmans, Goedbloed, and
  Kerner]{huysmans1991cp90}
G.T.A. Huysmans, J.P. Goedbloed, and W.~Kerner.
\newblock Cp90 conf. on comp.
\newblock \emph{Physics (Singapore: Word Scientific)}, 371, 1991.

\bibitem[Mikhailovskii et~al.(1997)Mikhailovskii, Huysmans, Kerner, and
  Sharapov]{mikhailovskii1997optimization}
A.B. Mikhailovskii, G.T.A. Huysmans, W.O.K. Kerner, and S.E. Sharapov.
\newblock Optimization of computational {MHD} normal-mode analysis for
  tokamaks.
\newblock \emph{Plasma Physics Reports}, 23\penalty0 (10):\penalty0 844--857,
  1997.

\bibitem[Pinches et~al.(2015)Pinches, Chapman, Lauber, Oliver, Sharapov,
  Shinohara, and Tani]{pinches2015energetic}
S.D. Pinches, I.T. Chapman, Ph.W. Lauber, H.J.C. Oliver, S.E. Sharapov,
  K.~Shinohara, and K.~Tani.
\newblock Energetic ions in {ITER} plasmas.
\newblock \emph{Physics of Plasmas}, 22\penalty0 (2):\penalty0 021807, 2015.

\bibitem[Borba and Kerner(1999)]{borba1999castor}
D.~Borba and W.~Kerner.
\newblock {CASTOR}-{K}: stability analysis of {A}lfv{\'e}n eigenmodes in the
  presence of energetic ions in tokamaks.
\newblock \emph{Journal of Computational Physics}, 153\penalty0 (1):\penalty0
  101--138, 1999.

\bibitem[Nabais et~al.(2015)Nabais, Borba, Coelho, Figueiredo, Ferreira,
  Loureiro, and Rodrigues]{nabais2015castor}
F.~Nabais, D.~Borba, R.~Coelho, A.~Figueiredo, J.~Ferreira, N.F. Loureiro, and
  P.~Rodrigues.
\newblock The {CASTOR}-{K} code, recent developments and applications.
\newblock \emph{Plasma Science and Technology}, 17\penalty0 (2):\penalty0 89,
  2015.

\bibitem[Nabais et~al.(2018)Nabais, Aslanyan, Borba, Coelho, Dumont, Ferreira,
  Figueiredo, Fitzgerald, Lerche, Mailloux, et~al.]{nabais2018tae}
F.~Nabais, V.~Aslanyan, D.~Borba, R.~Coelho, R.~Dumont, J.~Ferreira,
  A.~Figueiredo, M.~Fitzgerald, E.~Lerche, J.~Mailloux, et~al.
\newblock {TAE} stability calculations compared to {TAE} antenna results in
  {JET}.
\newblock \emph{Nuclear Fusion}, 58\penalty0 (8):\penalty0 082007, 2018.

\bibitem[Lauber(2015)]{lauber2015local}
P.~Lauber.
\newblock Local and global kinetic stability analysis of alfv{\'e}n eigenmodes
  in the 15 {MA} {ITER} scenario.
\newblock \emph{Plasma Physics and Controlled Fusion}, 57\penalty0
  (5):\penalty0 054011, 2015.

\bibitem[Holod et~al.(2009)Holod, Zhang, Xiao, and
  Lin]{holod2009electromagnetic}
I.~Holod, W.L. Zhang, Y.~Xiao, and Z.~Lin.
\newblock Electromagnetic formulation of global gyrokinetic particle simulation
  in toroidal geometry.
\newblock \emph{Physics of Plasmas}, 16\penalty0 (12):\penalty0 122307, 2009.

\bibitem[Hubbard et~al.(2017)Hubbard, Baek, Brunner, Creely, Cziegler, Edlund,
  Hughes, LaBombard, Lin, Liu, et~al.]{hubbard2017physics}
A.E. Hubbard, S.G. Baek, D.~Brunner, A.J. Creely, I.~Cziegler, E.~Edlund, J.W.
  Hughes, B.~LaBombard, Y.~Lin, Z.~Liu, et~al.
\newblock Physics and performance of the {I}-mode regime over an expanded
  operating space on {A}lcator {C}-{M}od.
\newblock \emph{Nuclear Fusion}, 57\penalty0 (12):\penalty0 126039, 2017.

\bibitem[Pace et~al.(2018)Pace, Austin, Bardoczi, Collins, Crowley, Davis, Du,
  Ferron, Grierson, Heidbrink, et~al.]{pace2018dynamic}
D.C. Pace, M.E. Austin, L.~Bardoczi, C.S. Collins, B.~Crowley, E.~Davis, X.~Du,
  J.~Ferron, B.A. Grierson, W.W. Heidbrink, et~al.
\newblock Dynamic neutral beam current and voltage control to improve beam
  efficacy in tokamaks.
\newblock \emph{Physics of Plasmas}, 25\penalty0 (5):\penalty0 056109, 2018.

\bibitem[Wong et~al.(1991)Wong, Fonck, Paul, Roberts, Fredrickson, Nazikian,
  Park, Bell, Bretz, Budny, et~al.]{wong1991excitation}
K.L. Wong, R.J. Fonck, SF~Paul, D.R. Roberts, E.D. Fredrickson, R.~Nazikian,
  H.K. Park, M.~Bell, N.L. Bretz, R.~Budny, et~al.
\newblock Excitation of toroidal alfv{\'e}n eigenmodes in tftr.
\newblock \emph{Physical Review Letters}, 66\penalty0 (14):\penalty0 1874,
  1991.

\bibitem[Gorelenkov et~al.(1999)Gorelenkov, Cheng, and Fu]{gorelenkov1999fast}
N.N. Gorelenkov, C.Z. Cheng, and G.Y. Fu.
\newblock Fast particle finite orbit width and {L}armor radius effects on low-n
  toroidicity induced {A}lfv{\'e}n eigenmode excitation.
\newblock \emph{Physics of Plasmas}, 6\penalty0 (7):\penalty0 2802--2807, 1999.

\bibitem[Kuang et~al.(2018)Kuang, Cao, Creely, Dennett, Hecla, LaBombard,
  Tinguely, Tolman, Hoffman, Major, et~al.]{fed}
A.Q. Kuang, N.M. Cao, A.J. Creely, C.A. Dennett, J.~Hecla, B.~LaBombard, R.A.
  Tinguely, E.A. Tolman, H.~Hoffman, M.~Major, et~al.
\newblock Conceptual design study for heat exhaust management in the {ARC}
  fusion pilot plant.
\newblock \emph{Fusion Engineering and Design}, 137:\penalty0 221--242, 2018.

\bibitem[Greenwald et~al.(2018{\natexlab{b}})Greenwald, Brunner, Hartwig, Irby,
  LaBombard, Lin, Marmar, Mumgaard, Sorbom, White,
  et~al.]{greenwald2018performance}
M.~Greenwald, D.~Brunner, Z.~Hartwig, J.~Irby, B.~LaBombard, Y.~Lin, E.~Marmar,
  R.~Mumgaard, B.~Sorbom, A.~White, et~al.
\newblock Performance projections for {SPARC}.
\newblock \emph{Bulletin of the American Physical Society}, 2018{\natexlab{b}}.

\bibitem[Fasoli et~al.(2007)Fasoli, Gormenzano, Berk, Breizman, Briguglio,
  Darrow, Gorelenkov, Heidbrink, Jaun, Konovalov, Nazikian, Noterdaeme,
  Sharapov, Shinohara, Testa, Tobita, Todo, Vlad, and
  Zonca]{0029-5515-47-6-S05}
A.~Fasoli, C.~Gormenzano, H.L. Berk, B.~Breizman, S.~Briguglio, D.S. Darrow,
  N.~Gorelenkov, W.W. Heidbrink, A.~Jaun, S.V. Konovalov, R.~Nazikian, J.-M.
  Noterdaeme, S.~Sharapov, K.~Shinohara, D.~Testa, K.~Tobita, Y.~Todo, G.~Vlad,
  and F.~Zonca.
\newblock Progress in the {ITER} {P}hysics {B}asis {C}hapter 5: Physics of
  energetic ions.
\newblock \emph{Nuclear Fusion}, 47\penalty0 (6):\penalty0 S264, 2007.

\bibitem[Vlad et~al.(2004)Vlad, Briguglio, Fogaccia, and
  Zonca]{vlad2004consistency}
G~Vlad, S~Briguglio, G~Fogaccia, and F~Zonca.
\newblock Consistency of proposed burning plasma scenarios with alpha-particle
  transport induced by alfv{\'e}nic instabilities.
\newblock \emph{Plasma Physics and Controlled Fusion}, 46\penalty0
  (7):\penalty0 S81, 2004.

\bibitem[Zonca et~al.(1996)Zonca, Chen, and Santoro]{zonca1996kinetic}
F.~Zonca, L.~Chen, and R.A. Santoro.
\newblock Kinetic theory of low-frequency alfv{\'e}n modes in tokamaks.
\newblock \emph{Plasma Physics and Controlled Fusion}, 38\penalty0
  (11):\penalty0 2011, 1996.

\bibitem[Ma et~al.(2014)Ma, Chavdarovski, Ye, and Wang]{ma2014linear}
R.~Ma, I.~Chavdarovski, G.~Ye, and X.~Wang.
\newblock Linear dispersion relation of beta-induced alfv{\'e}n eigenmodes in
  presence of anisotropic energetic ions.
\newblock \emph{Physics of Plasmas}, 21\penalty0 (6):\penalty0 062120, 2014.

\bibitem[Goldston and Rutherford(1995)]{goldston1995introduction}
R.J. Goldston and P.H. Rutherford.
\newblock \emph{Introduction to plasma physics}.
\newblock CRC Press, 1995.

\end{thebibliography}
\end{document}